\documentclass[12pt,preprint]{emulateapj}
\RequirePackage[normalem]{ulem} %DIF PREAMBLE
\RequirePackage{color}\definecolor{RED}{rgb}{1,0,0}\definecolor{BLUE}{rgb}{0,0,1} %DIF PREAMBLE
\slugcomment{Submitted 2013 December 25, Accepted 2014 June 17}

\shorttitle{\textit{Suzaku} Study of Type I Seyfert Galaxy NGC 3227}
\shortauthors{Noda et al.}

\begin{document}

%% LaTeX will automatically break titles if they run longer than
%% one line. However, you may use \\ to force a line break if
%% you desire.

\title{\textit{Suzaku} Studies of the Central Engine in the Typical Type I Seyfert  NGC 3227: 
Detection of Multiple Primary X-ray Continua with Distinct Properties}

%% Use \author, \affil, and the \and command to format
%% author and affiliation information.
%% Note that \email has replaced the old \authoremail command
%% from AASTeX v4.0. You can use \email to mark an email address
%% anywhere in the paper, not just in the front matter.
%% As in the title, use \\ to force line breaks.

\author{Hirofumi Noda\altaffilmark{1}, Kazuo Makishima\altaffilmark{2,3,4}, 
Shin'ya Yamada\altaffilmark{5}, Kazuhiro Nakazawa\altaffilmark{2}, \\
Soki Sakurai\altaffilmark{2}, and Katsuma Miyake\altaffilmark{2}}
%% Notice that each of these authors has alternate affiliations, which
%% are identified by the \altaffilmark after each name.  Specify alternate
%% affiliation information with \altaffiltext, with one command per each
%% affiliation.

%% Mark off your abstract in the ``abstract'' environment. In the manuscript
%% style, abstract will output a Received/Accepted line after the
%% title and affiliation information. No date will appear since the author
%% does not have this information. The dates will be filled in by the
%% editorial office after submission.

%----------------------------------------abstract--------------------------------------------------------
\begin{abstract}
The type I Seyfert galaxy NGC 3227 was observed by \textit{Suzaku} six times in 2008, 
with intervals of $\sim1$ week and net exposures of $\sim50$~ksec each.  Among the six 
observations, the source varied by nearly an order of magnitude, being brightest in the 1st 
observation with a 2--10~keV luminosity of $1.2\times10^{42}$~erg~s$^{-1}$, while faintest 
in the 4th with $2.9\times10^{41}$~erg~s$^{-1}$. As it became fainter, the continuum in a 
2--45~keV band became harder, while a narrow Fe-K$\alpha$ emission line, detected on all
 occasions at 6.4~keV of the source rest frame, remained approximately constant in the 
 photon flux. Through a method of variability-assisted broad-band spectroscopy 
 (e.g., Noda et al. 2013), the 2--45~keV spectrum of NGC 3227 was decomposed into three distinct
  components. One is a relatively soft power-law continuum with a photon index of $\sim 2.3$, 
  weakly absorbed and highly variable on time scales of $\sim5$~ksec; it was observed only 
  when the source was above a threshold luminosity of $\sim6.6\times10^{41}$~erg~s$^{-1}$ 
  (in 2--10~keV), and was responsible for further source brightening beyond. Another is a harder 
  and more absorbed continuum with a photon index of $\sim 1.6$, which persisted through the 
  six observations and varied slowly on time scales of a few weeks by a factor of $\sim2$.  
  This component, carrying a major fraction of the broad-band emission when the source is 
  below the threshold luminosity, is considered as an additional primary emission. The last 
  one is a reflection component with the narrow iron line, produced at large distances from 
  the central black hole.

\end{abstract}
%--------------------------------------------------------------------------------------------------------------

\altaffiltext{1}{Nishina Center, RIKEN (The institute of Physical and Chemical Research), 2-1, Hirosawa, Wako, Saitama 351-0198, Japan}
\altaffiltext{2}{Department of Physics, School of Science, The University of Tokyo, 7-3-1, Hongo, Bunkyo-ku, Tokyo 113-0033, Japan}
\altaffiltext{3}{MAXI team, RIKEN, 2-1, Hirosawa, Wako, Saitama 351-0198, Japan}
\altaffiltext{4}{Research Center for the Early Universe, School of Science, The University of Tokyo, 7-3-1, Hongo, Bunkyo-ku, Tokyo 113-0033, Japan}
\altaffiltext{5}{Department of Physics, Tokyo Metropolitan University, 1-1 Minami-Osawa, Hachioji, Tokyo, 192-0397 Japan}

\section{Introduction}
%---------------------------------------------------------

An X-ray spectrum from type I Active Galactic Nuclei (AGNs) generally 
exhibits a broad-band continuum extending up to several hundreds keV, 
with some fine structures including a prominent Fe-K$\alpha$ emission 
line at $\sim 6.4$ keV and an Fe-K absorption edge at $\sim 7.1$ keV.  
A significant fraction of the broad-band continuum is thought to be 
generated via inverse Compton scattering 
in a ``corona'' formed near a Super Massive Black Hole (SMBH) in the AGN. 
In previous studies, this broad-band thermal Comptonization continuum has often been 
regarded as a single Power-Law (PL) component which is generated in 
a single-zone corona with uniform physical parameters. 
As a result, excess signals above the single PL, frequently observed both in 
soft (below 3 keV) and hard (above 15 keV) X-ray ranges, have been 
interpreted as entities different from the primary Comptonization continuum  
(e.g., Risaliti \& Elvis 2004; Fabian \& Miniutti 2005), 
such as signals reprocessed by materials surrounding the SMBH.

By a wide-band X-ray timing analysis called Count-Count Correlation with Positive Offset 
(C3PO) method (Noda et al. 2011b, 2013a), which extends an initial attempt by Churazov et al. (2001), 
and an independent multi-wavelength spectral analysis 
(e.g., Mehdipour et al. 2011; Jin et al. 2012), 
the origin of the soft X-ray excess structure in several disk-dominated AGNs, 
including Mrk 509 in particular, has recently been clarified; 
it is  a thermal Comptonization component produced in an optically-thick corona, 
which is different from that emitting the broad-band PL continuum. 
These results imply that the central engine of AGNs consists of 
multiple regions with different physical parameters, generating several 
thermal Comptonization emissions with different spectral shapes 
and different variability characteristics. 
In contrast, 
the nature of the hard X-ray hump is still under discussion; 
some fraction of this feature is clearly due to reflection by neutral materials located 
far away from the SMBH,  
while the rest could be attributed to separate origins 
including relativistically-smeared reflection 
(e.g., Miniutti et al. 2007;  Walton et al. 2013; Risaliti et al. 2013) 
and/or a partially absorbed primary PL component 
(e.g., Miller et al. 2008; Miyakawa et al. 2012; Miller \& Turner 2013). 
As revealed by Noda et al. (2011a), 
even some part of the hard X-ray hump could be considered as primary emission.

To identify the origin of the hard X-ray hump, in Noda et al. (2013b) 
we applied the above mentioned C3PO method 
to 3--45 keV \textit{Suzaku} data of the bright Seyfert 1 galaxy NGC 3516, 
after previous works by, e.g., Taylar et al. (2003) and Vaughan \& Fabian (2004). 
As a result, we found that some fraction of the hard X-ray hump of NGC 3516 is due,  
as expected, to neutral and non-smeared reflection, 
while the rest is attributed to a second PL component  with a significantly flatter spectral shape, 
varying more slowly than the long-known broad-band PL component. 
We further concluded in Noda et al. (2013b) that the new PL is also thermal Comptonization, 
and is produced in regions around the SMBH that are different from those producing 
the broad-band PL component, because these two continua have 
different spectral and timing properties. 
Thus, some fraction of the hard X-ray hump is also considered as the primary 
emission from the SMBH as originally pointed out by Noda et al. (2011a), 
rather than reprocessed or partially absorbed signals. 
Combined with the interpretation of the soft excess, we hence arrive at a 
novel view that the central engine of an AGN consists of at least three separate regions 
which boost soft photons (most likely from the accretion disk) 
via Comptonization into broad-band X-ray photons. 

Although we found (Noda et al. 2013b) 
the new hard PL component of NGC 3516 to be varying significantly on long time scales,  
we did not catch a moment in which it just varied.  
To catch that moment, we should select another AGN, 
which was observed by \textit{Suzaku} 
more frequently and exhibited larger intensity variations. 
Thus, in the present paper, we focus on the bright and well studied 
Seyfert 1 galaxy NGC 3227, 
which was observed 6 times by \textit{Suzaku},  
with intervals of about a week. 
This AGN has a redshift of $z=0.00386$ and a column density of the Galactic absorption 
$N_{\rm H} \sim 2 \times 10^{20}$ cm$^{-2}$. Its black hole mass and the typical 0.5--2 keV 
luminosity are $\sim 4 \times 10^7$ solar masses (e.g., Peterson et al. 2004) and 
$\sim 5 \times 10^{41}$ erg s$^{-1}$ (George et al. 1998), respectively, 
which imply a relatively low Eddington ratio around $\sim 0.001$, 
assuming a bolometric correction factor of $\sim 10$. 
Unless otherwise stated, 
errors in the present paper refer to 90\% confidence limits.

%=================
\section{Observation and Data Reduction}
%=================

%%%%%%%%%%%table 1%%%%%%%%%%%%

\begin{table}[b]
 \caption{Information on the \textit{Suzaku} observations of NGC 3227. }
 \label{all_tbl}
 \begin{center}
  \begin{tabular}{ccc}
   \hline\hline
 Observation start time &Observation ID & exposure (sec)   \\

   \hline
2008/10/28 08:12:52 	    & 703022010 & 58917.2  \\
2008/11/04 03:36:31    & 703022020 & 53699.5  \\
2008/11/12 02:48:55    & 703022030  &  56571.5 \\
 2008/11/20 17:00:00  & 703022040  & 64567.9 \\
 2008/11/27 21:29:20   & 703022050 & 79429.8  \\
 2008/12/02 14:28:03   & 703022060 & 51410.5  \\ \hline
        
  \end{tabular}
 \end{center} 
          
\end{table}
%%%%%%%%%%%table 1%%%%%%%%%%%%

As summarized in Table 1, NGC 3227 was observed with \textit{Suzaku} six times 
from October 28 to December 2 in 2008, with intervals of $\sim 1$ week, 
and the individual gross and net exposure were $\sim100$ ksec 
and $\sim 50$--70 ksec, respectively. 
In these observations, 
the XIS and HXD onboard \textit{Suzaku} were operated in their normal modes, 
and the source was placed at the XIS nominal position. 
The XIS and HXD-PIN data from these observations utilized here
were prepared via version 2.2 processing. 
In the present paper, we do not utilize the HXD-GSO data.

Below, soft X-ray signals are taken from XIS0 and XIS3 (collectively called XIS FI), 
while XIS1 is not utilized because of its relatively high background. 
On-source XIS FI events were accumulated over a circular region 
of $120''$ radius centered on the source, while the corresponding 
background was taken from an annular region  
with the inner and outer radii of  $180''$ and $270''$, respectively. 
The response matrices and ancillary response files were created 
by \texttt{xisrmfgen} and \texttt{xissimarfgen} (Ishisaki et al. 2007), respectively. 

We also analyze the HXD-PIN events prepared via version 2.2 processing 
in a similar way to those of the XIS. 
Non X-ray background included in the on-source HXD-PIN data 
was estimated by utilizing fake events created 
by a standard background model (Fukazawa et al. 2009), 
and contribution from the Cosmic X-ray background (Boldt et al. 1987) 
was calculated based on 
the spectral brightness model, 
$9.0 \times 10 ^{-9} (E/3~\mathrm{keV})^{-0.29} \exp (-E/40~\mathrm{keV})$
erg cm$^{-2}$ s$^{-1}$ str $^{-1}$ keV$^{-1}$ (Gruber et al. 1999). 
They were subtracted from the on-source data.

Another X-ray source NGC 3226, which is a low-luminosity AGN or a LINER, 
is located at $2'$ away from the center of NGC 3227, 
and its signals can contaminate both the extracted XIS FI and HXD-PIN events. 
It is actually visible in our XIS images. 
However, even in the 4th observation when NGC 3227 was faintest, 
the XIS count rate (including background) from a $20''$-radius circle 
centered on this AGN was $\gtrsim 10$ times higher than that from a region of 
the same radius centered on NGC 3226. 
 Therefore, our results are not significantly affected by this additional source.

%=====================3Í==========================
\section{Spectra and Light Curves}
%==================================================

%-----------------------------------3.1Í----------------------------------------------
\subsection{Time-averaged spectra}
%-----------------------------------------------------------------------------------------

%%%%%%%%%%%%%%%%figure1%%%%%%%%%%%%%%%%%%%%%%%%
\begin{figure}[b]
\epsscale{1.1}
\plotone{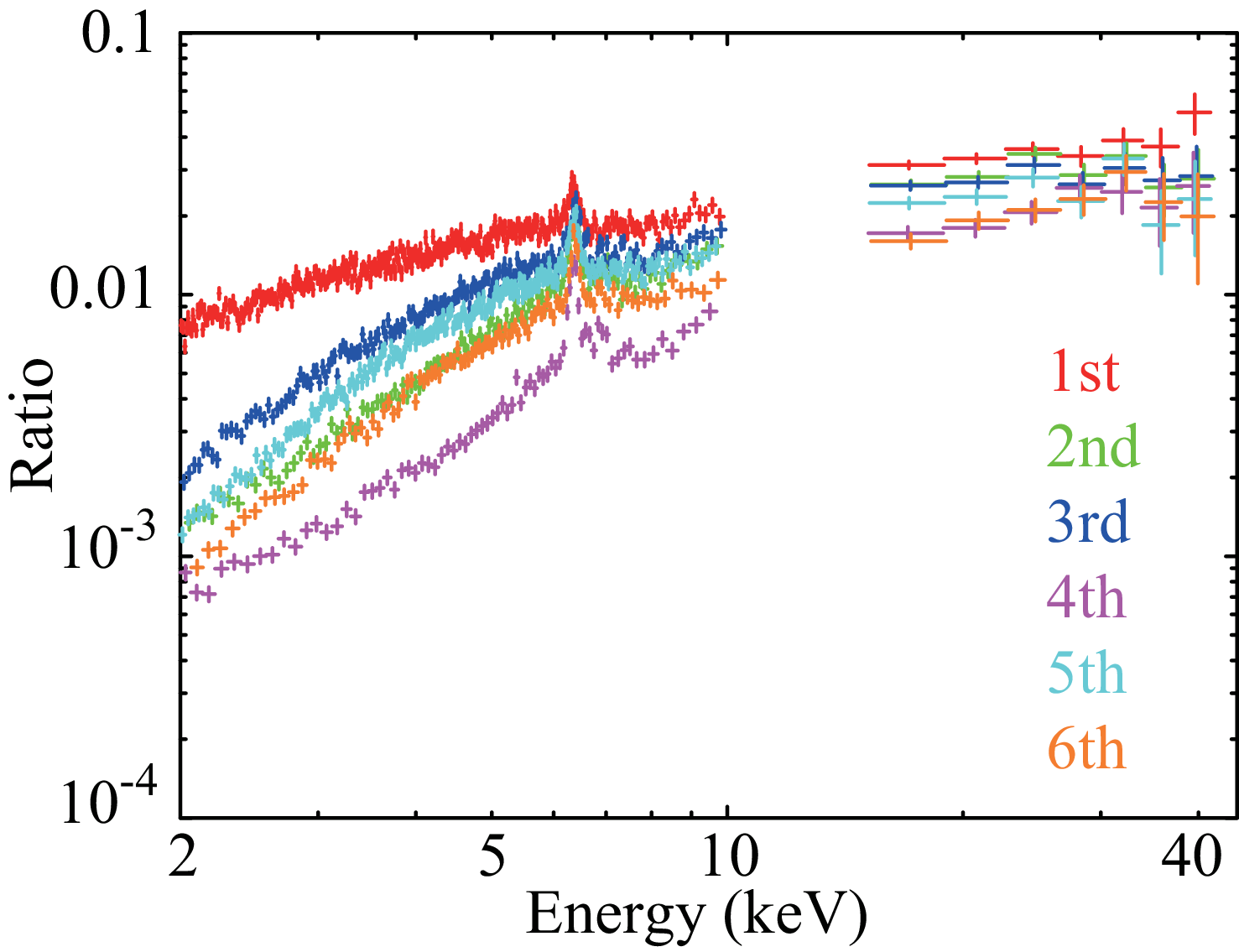}
\caption{Background-subtracted and time-averaged spectra of NGC 3227 obtained in 
the 1st (red), 2nd (green), 3rd (blue), 4th (purple), 5th (cyan), and 6th (orange) 
\textit{Suzaku} observations. 
They are shown in a form of ratios to a PL of which the photon index is 2.0 and 
the normalization is common (equivalent to a $\nu F_{\nu}$ form except that the 
deconvolution is not performed). 
The data below 10 keV are from the XIS FI, and those above 15 keV from HXD-PIN.}
\end{figure}
%%%%%%%%%%%%%%%%figure1%%%%%%%%%%%%%%%%%%%%%%%%

Figure 1 shows six time-averaged 2--45 keV \textit{Suzaku} spectra of NGC 3227, 
visualizing spectral changes among the 6 observations. 
They commonly exhibit a prominent Fe-K$\alpha$ emission line at $\sim6.35$ keV 
(6.4 keV in the rest frame), of which the intensity is apparently similar among the 6 data sets. 
In the faintest 4th observation, 
a clear Fe-K absorption edge at $\sim7.1$ keV and a Ni-K$\alpha$ emission 
line at $\sim7.15$ keV also appeared. 
None of these spectra show noticeable evidence for warm absorber, 
which would produce deep and narrow absorption lines in the XIS data. 

%%%%%%%%%%%%%%%%figure2%%%%%%%%%%%%%%%%%%%%%%%%
\begin{figure*}[t]
\epsscale{1}
\plotone{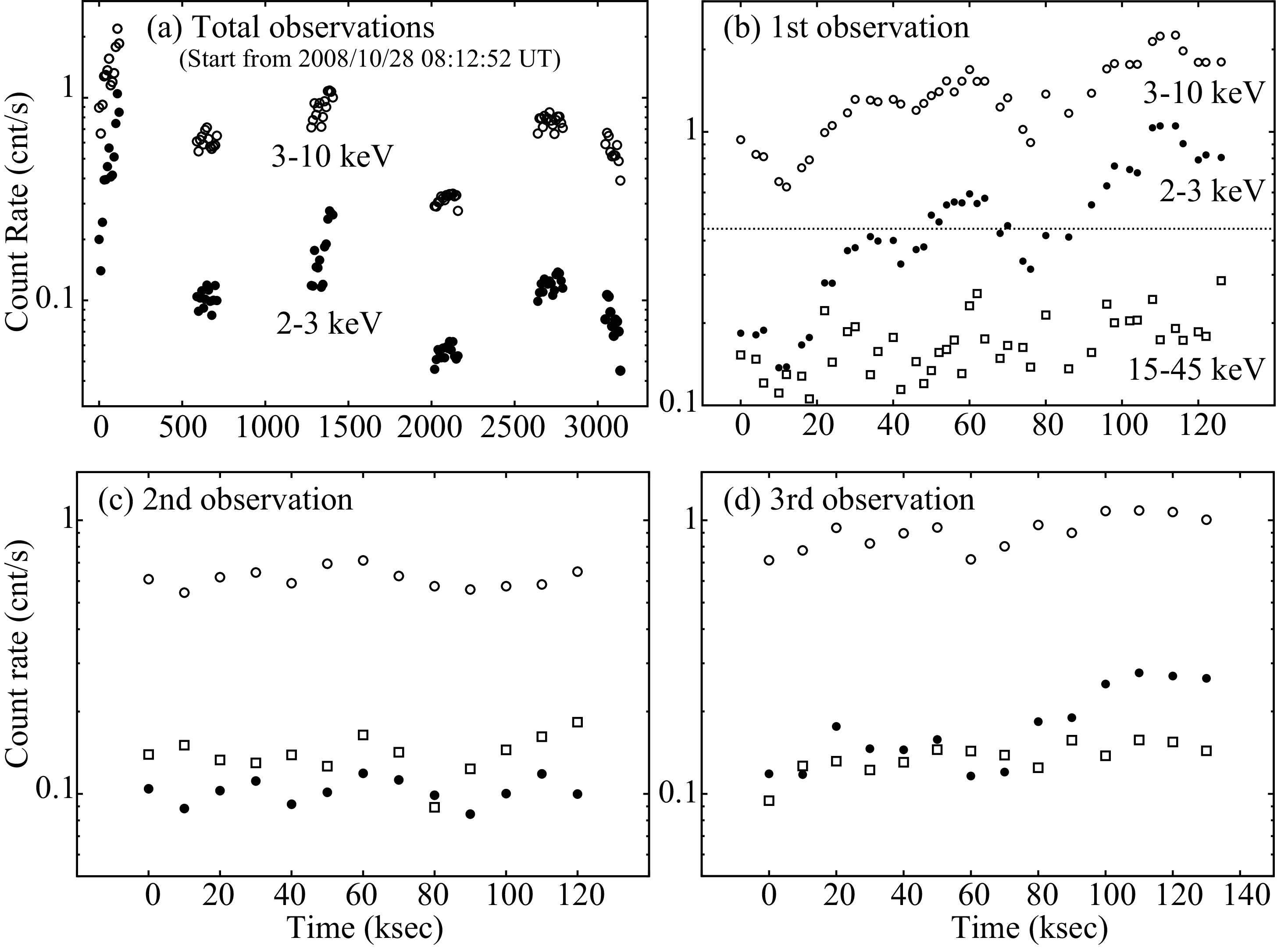}
\caption{Panel (a) shows 
background-subtracted and dead-time corrected 2--3 keV (filled) and 3--10 keV (open)
light curves recorded in the 6 observations of NGC 3227 with 10 ksec binning. 
Panel (b), (c), and (d) show expanded light curves in the 1st, 2nd, and the 3rd observations, respectively. 
The plots in panel (b) have a bin size of 2 ksec, while the others 10 ksec. 
In these panels, filled and open circles are the same XIS FI data as in panel (a), 
while open squares represent 15--45 keV background-subtracted HXD count rate. 
Statistical errors associated with these data points are all less than $\sim 0.03$ cnt s$^{-1}$, 
and are hence omitted. The HXD-PIN data points, however, are subject to systematic errors 
due to background uncertainty, which is typically $\sim0.02$ cnt s$^{-1}$.  }
\end{figure*}
%%%%%%%%%%%%%%%%figure2%%%%%%%%%%%%%%%%%%%%%%%%

One major difference among the 6 time-averaged spectra appears 
in their continuum spectral shapes. 
The continuum in the 1st observation,  
when the source was brightest and most variable on short time scales (see \S3.2), 
is steeper (in terms of photon index) than those in the other observations. 
In 2--10 keV, the spectral shape and brightness  of the 1st observation 
are similar to those obtained with \textit{XMM-Newton} (Markowitz et al. 2009). 
On the other hand,  the continuum in the 4th observation, the faintest data, is rather hard. 
Thus, the source variation among the 6 observations is much larger in 
lower energies than in a range above $\sim 15$ keV. 
In addition, the spectrum in the 4th observation is concave below $\sim5$ keV, 
while those in the other observations are convex.

%-------------------------------3.2Í---------------------------------------------------
\subsection{Light curves}
%-----------------------------------------------------------------------------------------

%%%%%%%%%%%%%%%%figure3%%%%%%%%%%%%%%%%%%%%%%%%
\begin{figure}[b]
\epsscale{1.1}
\plotone{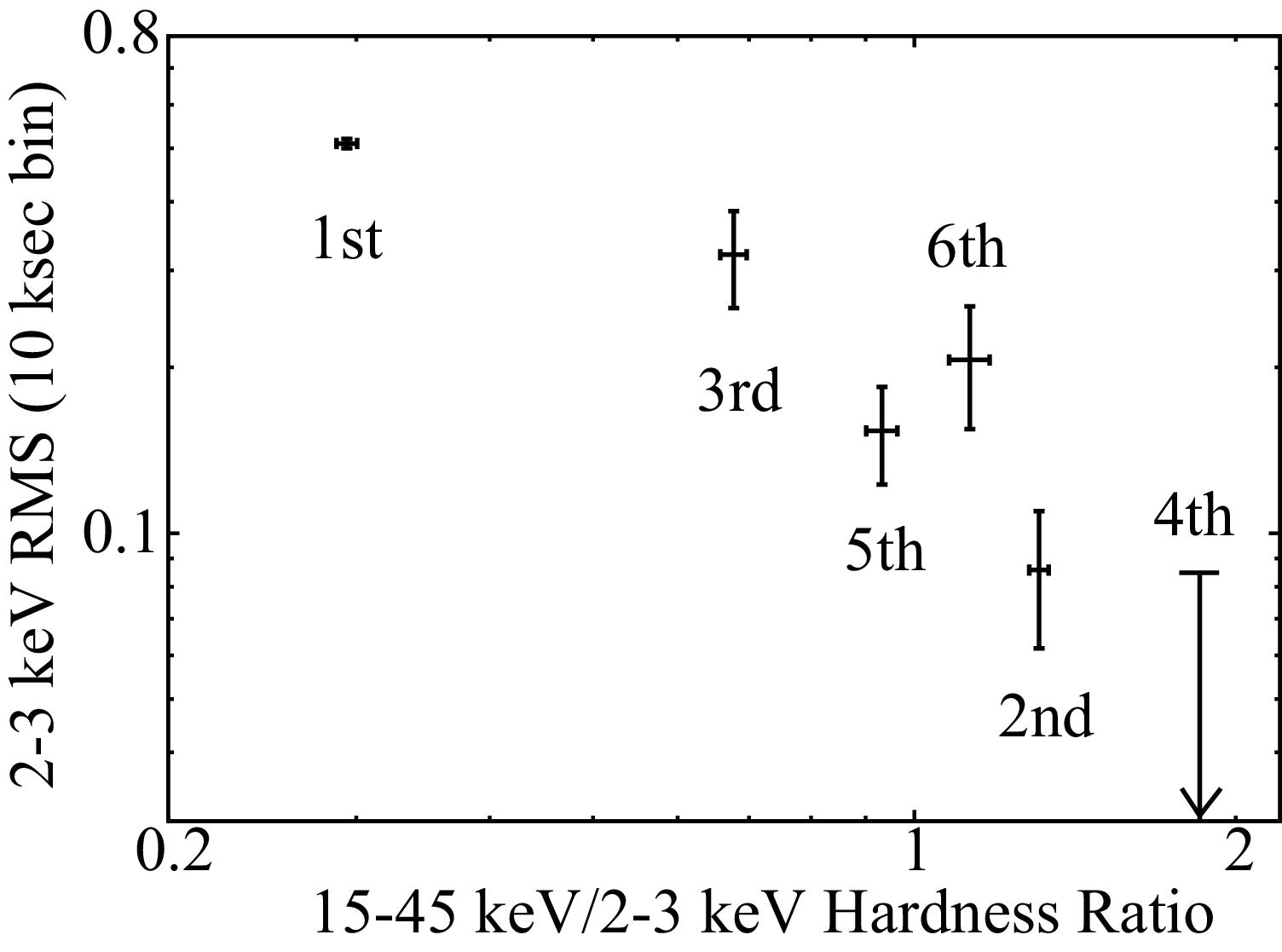}
\caption{A scatter plot between hardness ratio (15--45 vs. 2--3 keV band) of NGC 3227, 
and the 2--3 keV RMS variation calculated with a 10 ksec binning. }
\end{figure}
%%%%%%%%%%%%%%%%figure3%%%%%%%%%%%%%%%%%%%%%

%%%%%%%%%%%%%%%%%figure4%%%%%%%%%%%%%%%%%%%%%%%
\begin{figure*}[t]
\epsscale{1}
\plotone{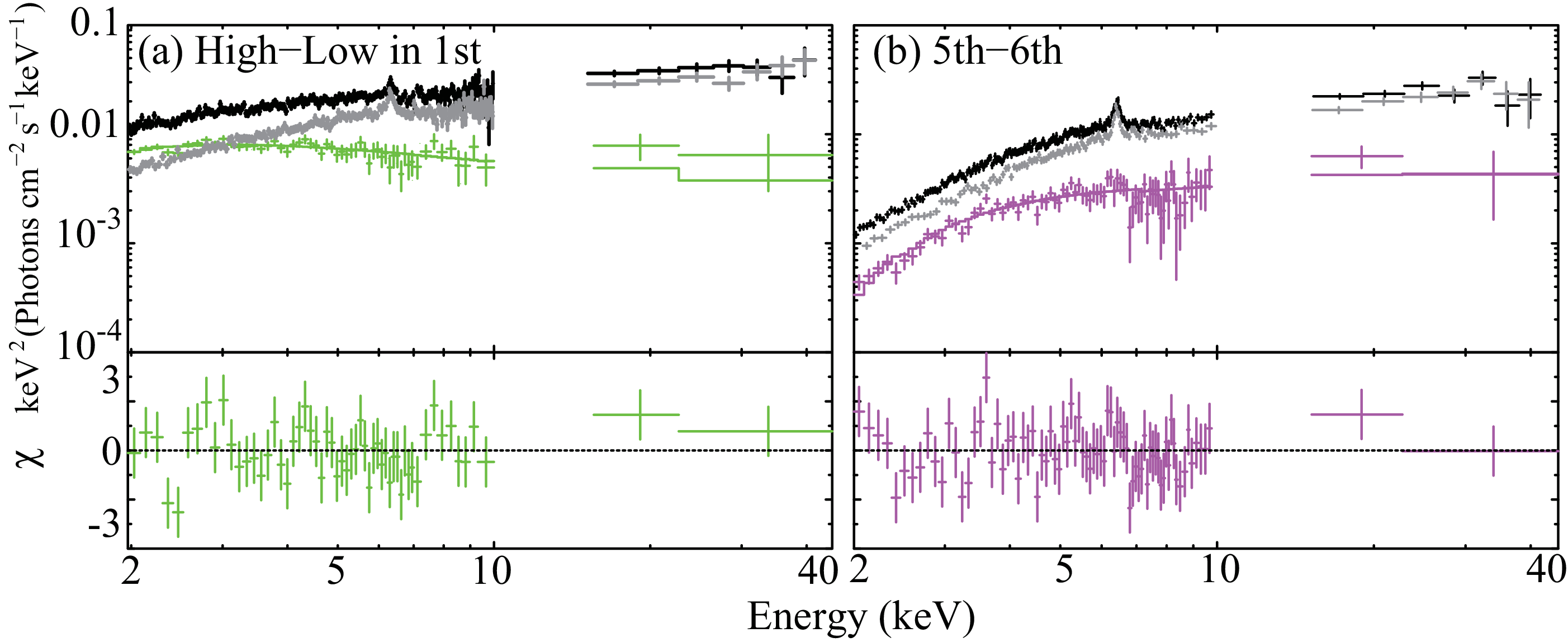}
\caption{(a) A short-term difference spectrum (green) created 
by subtracting the Low- (grey) from the High-phase spectrum 
(black) in the 1st observation. (b) A long-term difference spectrum (purple), 
extracted by subtracting a time-averaged spectrum 
in the 6th observation (grey) from that in the 5th one (black). }
\end{figure*}
%%%%%%%%%%%%%%%%%figure4%%%%%%%%%%%%%%%%%%%%%

Figure 2 shows 2--3 keV, 3--10 keV, and 15--45 keV light curves of all the observations. 
Thus, the source was more variable in the 1st and 3rd observations 
than in the others. 
Especially in the 1st observation, 
the intensity varied by about an order of magnitude within $\sim 100$ ksec, 
and the shortest peak-to-peak variation timescale was $\sim 5$ ksec which is 
equivalent to a light travel time across $\lesssim 25~R_{\rm g}/c$, 
where $R_{\rm g} = GM_{\rm BH}/c^2$ 
is the gravitational radius for a SMBH mass of $M_{\rm BH} = 4\times10^7~M_{\odot}$ (\S1), 
while $G$, $c$, and $M_{\odot}$ denote the gravitational constant, 
the light velocity, and  the solar mass, respectively. In contrast, 
the other observations, i.e., the 2nd, 4th, 5th, and 6th, individually 
showed considerably smaller short-term (i.e., intra-data-set) variations; 
$\sim50$\% for the 6th data set, and less than 20\% for the 2nd, 4th, and 5th.
(Although the HXD-PIN variation amplitude is seen to be rather high in Fig. 2, 
it is due to background-subtraction uncertainty.)
Therefore, NGC 3227 exhibited both short- and long-term variations, 
with complicated spectral shape changes (Fig. 1). 

Considering the spectral behavior seen in Fig. 1 and the light curves in Fig. 2, 
we expect that the source is more variable in a soft X-ray range 
when the broad-band spectrum becomes softer. 
To confirm this prediction, 
we present in Fig. 3 a scatter plot between the hardness ratio, made by dividing the 15--45 keV 
time-averaged count rate by that in the 2--3 keV one, against the root mean square (RMS) variability 
calculated using the 2--3 keV light curves in Fig. 2(a). 
As expected, the scatter plot reveals a clear negative correlation. 
Thus, the broad-band X-ray emission is suggested to consist of a softer continuum 
which varies both on the short (several hours; Fig. 2 and 3) 
and long (several weeks; Fig. 1) time scales, 
and a harder one which varies only on the long time scale (Fig. 1). 
When the former increases, the source is considered to get softer on average and more variable, 
as represented by the 1st and 3rd observations. 

%%%%%%%%%%%Table 2%%%%%%%%%%
\renewcommand{\arraystretch}{1}
\begin{table}[b]
 \caption{Results of the fits to the difference spectra and the C3PO-derived variable spectra.$^{\rm a}$}
 \label{all_tbl}
 \begin{center}
  \begin{tabular}{ccccc}
   \hline\hline
    Component & Parameter  & Difference& C3PO variable\\
          \hline
     
       & &  \multicolumn{2}{c}{Short term (Bright branch)}\\[1.5ex]
                      	 
     \texttt{wabs} &$ N_{\rm H0}^{\rm b}$
			& $1.2 \pm 0.8$
			& $1.8^{+1.6}_{-1.5}$\\

  \texttt{cutoffPL} & $\Gamma_0$
  			& $2.42^{+0.20}_{-0.18}$
			&  $2.41^{+0.20}_{-0.18}$\\
			
			& $E_{\rm cut}$~(keV)
			&  \multicolumn{2}{c}{200 (fix)}\\
			
			&$N_{\rm PL0}^{\rm c}$
			&$1.56^{+0.06}_{-0.04}$
			&$2.01^{+1.00}_{-0.61}$\\[1.5ex]

   $\chi^{2}$/d.o.f. & &  51.8/50 & 4.75/13   \\\hline
   
  & &  \multicolumn{2}{c}{Long term (Faint branch)}\\[1.5ex]
   
        \texttt{wabs} &$ N_{\rm H1}^{\rm b}$
                          & $5.3 \pm 0.8$
                          &$8.4^{+2.1}_{-2.0}$\\

  \texttt{cutoffPL} & $\Gamma_1$
			& $1.84^{+0.19}_{-0.17}$
			&$1.61^{+0.21}_{-0.19}$\\
			
			& $E_{\rm cut}$~(keV)
			&  \multicolumn{2}{c}{200 (fix)}\\
			
			&$N_{\rm PL1}^{\rm c}$
			&$0.26^{+0.07}_{-0.10}$
			&$0.43^{+0.23}_{-0.14}$\\[1.5ex]

   $\chi^{2}$/d.o.f. &  & 80.1/78 & 8.3/13 \\

\hline\hline

  \end{tabular}
\end{center}
   	{\small
	$^{\rm a}$ The C3PO-derived variable spectra refer to the case with $C=0$ in eq. (2).\\
	$^{\rm b}$ Equivalent hydrogen column density in  $10^{22}$ cm$^{-2}$. \\
         $^{\rm c}$ The \texttt{cutoffPL} normalization at 1 keV, in units of $10^{-3}$~photons~keV$^{-1}$~cm$^{-2}$~s$^{-1}$~at 1 keV.\\
         
}

\end{table}

%%%%%%%%%%%Table 2%%%%%%%%%%%%

%======================4Í=======================
\section{Conventional and New Timing Analyses}
%================================================

%----------------------------------------4.1Í------------------------------------------
\subsection{Difference spectrum analyses}
%-------------------------------------------------------------------------------------------

To identify the spectral components responsible for the short- and long-term variations (\S3.2), 
the conventional difference spectrum analysis was first employed. 
For the short-term variation, 
we divided the 1st observation into a High- and Low-phase,  
in which the 2--3 keV count rate is above and below 
the average shown by a dotted line in Fig. 2(b), respectively, 
and extracted a spectrum from each phase. 
Subtracting the Low- from the High-phase spectrum, 
we obtained a short-term difference spectrum shown in green in Fig. 4(a). 
On the other hand, 
a long-term difference spectrum can be derived by 
subtracting  the time-averaged spectrum of the 6th observation (grey in Fig. 4b) 
from that of the 5th one (black in Fig. 4b). 
These two particular observations were utilized because they are 
likely to be less contributed by the variable softer continuum. 
The obtained long-term difference spectrum is shown in Fig. 4(b) in purple. 
While the short-term difference spectrum (green in Fig. 4a) 
is relatively soft as expected in \S3.2, the long-term  one (purple 
in Fig. 4b) is significantly harder, and appears more absorbed. 

The neutral Fe-K$\alpha$ emission line, attributable to cold reflection, 
did not vary significantly in its intensity, and such an iron line feature is not 
seen either in the short-term or long-term difference spectra obtained in 
this way. 
Therefore, the two difference spectra are both likely to be dominated by primary photons,  
although possible presence of relatively featureless secondary 
signals are not yet ruled out.  
Given this, we fitted them with an absorbed cutoff PL model (\texttt{wabs * cutoffPL} in XSPEC12), 
in which the column density, the photon index, and the normalization were left free, 
while the cutoff energy was fixed at 200 keV. 
As shown in Fig. 4 and Table 2, 
the fits to the short- and long-term difference spectra became both acceptable 
with $\chi^2$/d.o.f.=51.8/50 and 80.1/78, respectively. 
As expected, 
the photon index of the former, $\Gamma_0 \sim 2.4$, is significantly
larger than that of the latter, $\Gamma_1 \sim 1.8$. 
In addition, the column density of the absorption to the long-term one 
was confirmed to be significantly higher than that to the other.

Even when we employed other count-rate thresholds separating the High- and Low-phase 
in the 1st dataset, e.g., a higher value as $0.7$ cnt s$^{-1}$ or a lower value 
as $0.3$ cnt s$^{-1}$, the obtained photon index and column density of 
the short-term difference spectrum became, within errors, consistent with 
those shown in Table 2. 
Similarly, 
other combinations among the 2nd, 4th, 5th, and 6th data sets gave long-term difference 
spectra that are consistent, within errors, with that shown in Fig. 4(b). 
This implies that the absorption appearing on the long-term 
difference spectrum is a general property of these data sets, 
rather than specific to particular data sets chosen for the calculation.
Thus, the overall source behavior requires the presence of (at least) 
two separated continuum components with distinct spectral and variability characteristics.

As described so far, the conventional difference spectrum method 
allowed us to identify the two spectral components, 
the short-term variable component represented by a weakly-absorbed soft PL, 
and the long-term variable one explained by a highly-absorbed hard PL. 
However, their mutual relation is not yet clear; 
e.g., it is uncertain whether the purple spectrum in Fig. 4(b) is contributed to 
some extent by the green one in Fig. 4(a), and 
whether the green and purple variable continua 
co-exist or not. This urges us to proceed to a 
more sophisticated method, as described in \S4.2. 

%----------------------------------------4.2Í------------------------------------------
\subsection{Count-Count Plot}
%-------------------------------------------------------------------------------------------

To characterize the variability more systematically in the way as Noda et al. (2013b), 
we made in Fig. 5 a Count-Count Plot (CCP) between  2--3 keV vs. 3--10 keV bands, 
incorporating all the six data sets.  
Interestingly, the data points in the CCP are distributed along a single broken line 
with a clear break point at a 2--3 keV count rate of $x_{\rm B} \sim0.16$ cnt s$^{-1}$. 
Hereafter, we call the data distributions above and below the break point 
Bright and Faint branches, respectively. 
Most of the data points from the 1st (red) and 3rd (blue) observations, 
in which the source was bright and variable, are distributed on the Bright branch, 
while those from the others (green, cyan, orange, and purple) on the Faint branch. 

To quantify the data distributions in the two branches in Fig. 5, 
we individually fitted them with a linear function expressed by 
\begin{equation}
y = Ax + B,
\end{equation}
in which $x$ and $y$ represent abscissa and ordinate, respectively, while 
the slope $A$ and the offset $B$ were both left free. 
Like in Noda et al. (2013b), the regression in the fits were performed 
by the Bivariate Correlated Errors and intrinsic Scatter (BCES) 
algorithm (Akritas \& Bershady 1996) which 
is one of the most common methods to take into account errors in both $x$ and $y$. 
As a result, 
the Bright- and Faint-branch data distributions were reproduced with ($A$, $B$)
=($1.48\pm0.08$, $0.62 \pm 0.03$) and ($4.55\pm0.36$, $0.18\pm0.04$), respectively. 
These fits gave $\chi^2/$d.o.f.=40.0/25 (Bright) and 74.0/53 (Faint), 
calculated after including systematic errors of 5\% and 7\%, respectively, 
which are necessary at least to take into account vignetting effects due to satellite attitude changes and slight absorption variations among different observations. 

%%%%%%%%%%%%%%%%%%figure5%%%%%%%%%%%%%%%%%%%
\begin{figure}[t]
\epsscale{1.1}
\plotone{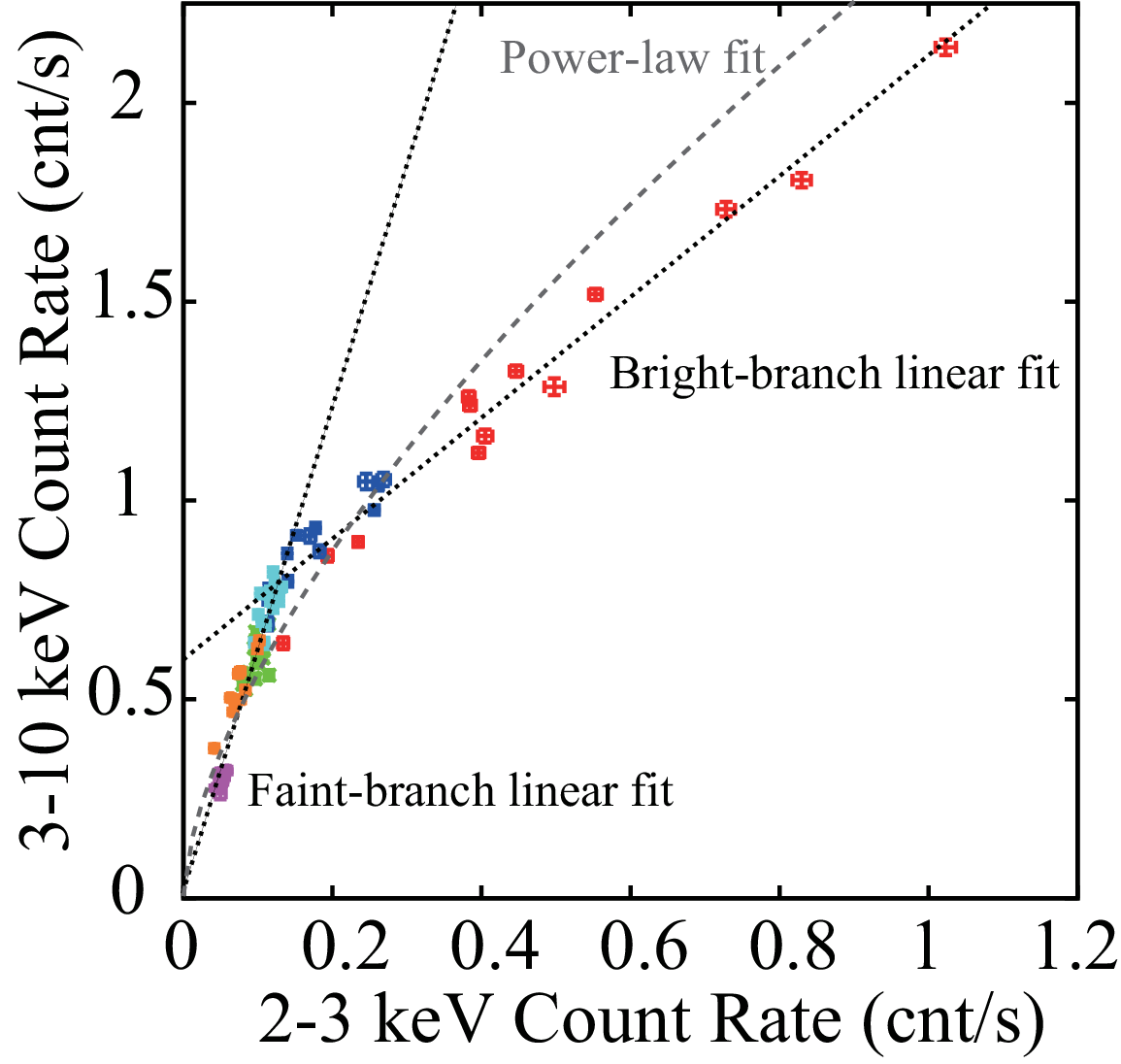}
\caption{A 2--3 keV vs. 3--10 keV CCP of NGC 3227, combining the entire six observations, 
with the same colors as in Fig. 1. All data are binned into 10 ksec, and the error bars 
refer to statistical $\pm1\sigma$ range. Black dotted lines represent linear fits 
to the Bright and Faint-branch data points, 
while grey dashed line the power-law fit to all the data points. }
\end{figure}
%%%%%%%%%%%%%%%%%%figure5%%%%%%%%%%%%%%%%%%%%

%%%%%%%%%%%%%%%%%%figure6%%%%%%%%%%%%%%%%%%%
\begin{figure*}[t]
\epsscale{1}
\plotone{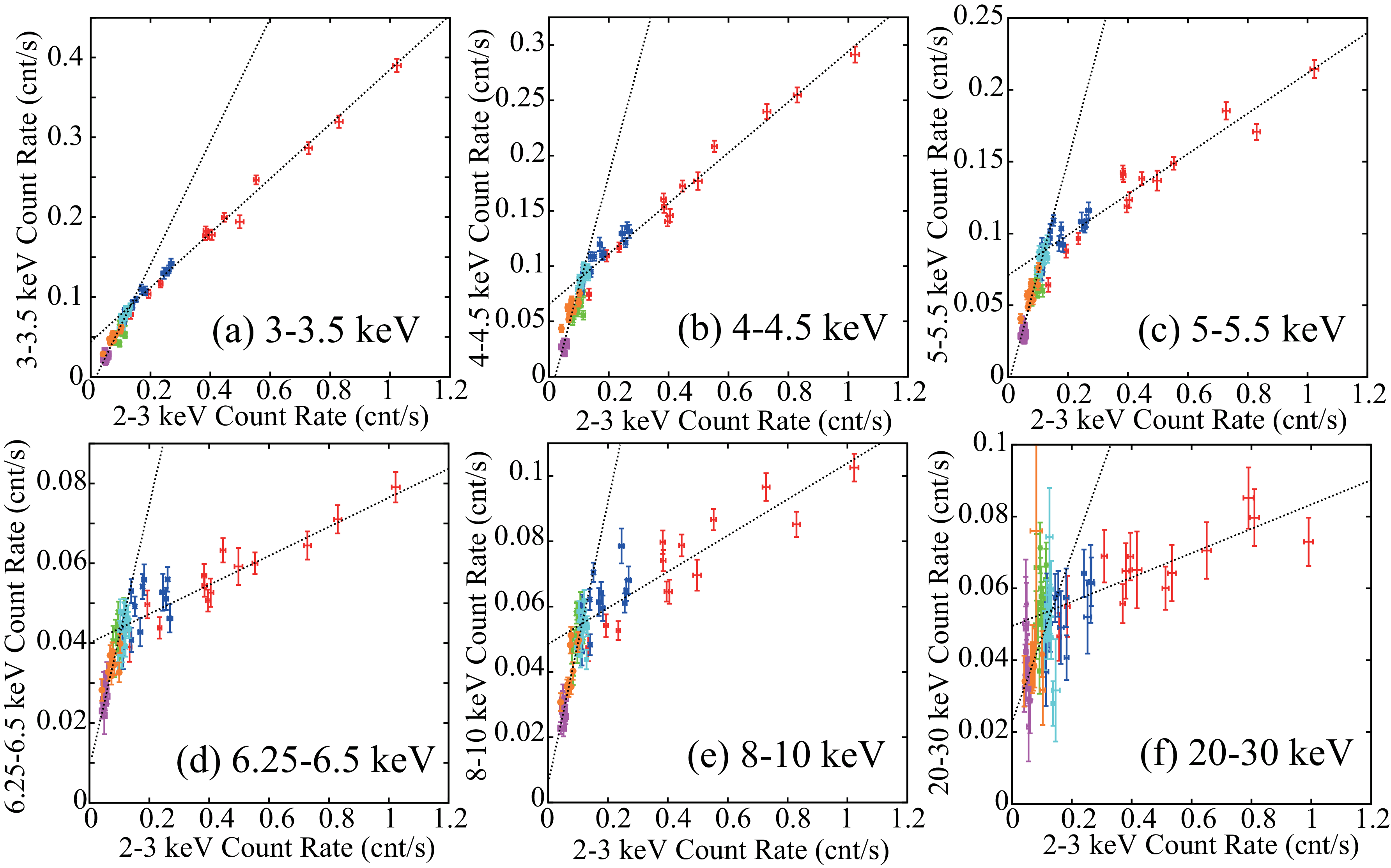}
\caption{Examples of the finer-band vs. 2--3 keV CCPs of NGC 3227, 
in which all the six observations are plotted together 
with the same colors as those in Fig. 5. 
The bin size is 10 ksec, and the error bars show statistical $\pm 1\sigma$ range. 
The two dotted lines represent the regression straight lines obtained 
separately  in the two branches. 
}
\end{figure*}
%%%%%%%%%%%%%%%%%%figure6%%%%%%%%%%%%%%%%%%%%

%%%%%%%%%%%table 3%%%%%%%%%%%%
\renewcommand{\arraystretch}{1}
\begin{table*}[t]
 \caption{Parameters of the linear fits to the Bright- and Faint-branches in the 16 CCPs, 
 representing the short- and long-term variations, respectively. }
 \label{ngc3227_ccp_fit}
 \small
 \begin{center}
  \begin{tabular}{ccccc}
   \hline\hline
    &\multicolumn{2}{c}{Short-term (Bright)} & \multicolumn{2}{c}{Long-term (Faint)}\\\hline
 Range (keV) & Slopes &Offsets$\times 10^2$ & Slopes &Offsets$\times 10^2$\\
   \hline
   3--3.5 & $0.34\pm0.01$& $4.60\pm0.32$ & $0.63\pm0.05$& $-0.54\pm 0.33 $ \\[-1ex]
    3.5--4   &$0.27\pm0.01$& $6.24\pm0.26$ & $0.66\pm0.05$& $-0.29\pm0.36$ \\[-1ex]
     4--4.5 & $0.22\pm0.01$& $6.92\pm0.41$  & $0.73\pm0.07$& $-0.54\pm0.44$ \\ [-1ex]
  4.5--5   & $0.17\pm0.01$ & $7.30\pm0.33$ & $0.78\pm0.06$& $-0.76\pm0.40$  \\[-1ex]
 5--5.5   & $0.14\pm0.01$ & $7.05\pm0.41$ & $0.69\pm0.06$& $-0.19\pm0.39$  \\[-1ex]
 5.5--6   & $0.10\pm0.01$ & $6.58\pm0.40$  & $0.63\pm0.05$& $0.15\pm0.33$  \\[-1ex]
 6--6.25    & $0.04\pm0.01$ & $3.43\pm0.18$  & $0.29\pm0.03$& $0.27\pm0.18$  \\[-1ex]
   6.25--6.5    & $0.03\pm0.01$ & $4.20\pm0.21$ & $0.27\pm0.02$& $1.35\pm0.17$  \\ [-1ex]
 6.5--6.75    & $0.03\pm0.01$ & $2.61\pm0.17$ & $0.22\pm0.02$& $0.39\pm0.15$ \\ [-1ex]
   6.75--7    &  $0.02\pm0.01$ & $2.13\pm0.20$ & $0.18\pm0.02$& $0.47\pm0.13$ \\ [-1ex]
 7--7.5    & $0.04\pm0.01$ & $3.58\pm0.20$ & $0.32\pm0.03$& $0.30\pm0.19$  \\ [-1ex]
  7.5--8    & $0.04\pm0.01$ & $2.17\pm0.21$ & $0.23\pm0.02$& $0.18\pm0.15$  \\ [-1ex]
  8--10    & $0.05\pm0.01$ & $4.99\pm0.32$ & $0.47\pm0.03$& $0.43\pm0.23$ \\ [-1ex]
15--20    & $0.04\pm0.01$ & $6.79\pm0.42$ & $0.27\pm0.09$& $3.56\pm0.67$  \\ [-1ex]
 20--30    & $0.04 \pm 0.01$ & $4.80\pm0.35$ & $0.23\pm0.08$& $2.72\pm0.57$ \\ [-1ex]
 30--45    & $0.00\pm0.01$ & $1.60\pm0.27$ & $0.09\pm0.06$& $0.58\pm0.42$ \\
	                            
      \hline\hline

  \end{tabular}
 \end{center}

\end{table*}
\renewcommand{\arraystretch}{1.0}
%%%%%%%%%%%table 3%%%%%%%%%%%%

To test an alternative interpretation of the CCP distribution, we also fitted it with a 
single power-law function, 
$y=Mx^N$ where $M$ and $N$ were left free (e.g., Uttley \& McHardy 2005). 
However, as shown in Fig. 5 in grey, the fit became significantly worse with $\chi^2$/d.o.f.=192.3/80, 
than the two straight line fit (114.0/78; sum of the Bright and Faint results).
Thus, the data distribution prefers the interpretation with a pair of straight lines with different slopes. 

As indicated by Fig. 2(b) and already noticed in \S3.2,  
the Bright-branch data, composed of the 1st and 3rd data sets, 
exhibit fast variations on a timescale of $<1$ day, 
while slower  ($\gtrsim 1$ week) variations are dominant in the Faint branch 
as represented by the behavior of the 2nd, 4th, 5th, and 6th data sets. 
The break count rate $x_{\rm B} \sim0.16$ cnt s$^{-1}$ is the point below which 
the short-term variability disappears, and 
the long-term one instead becomes dominant. 
Thus, the CCP data distribution 
provides evidence for the presence of independent short- and long-term variations, 
each caused presumably by intensity (but no spectral shape) changes of 
a characteristic spectral component. 
We further expect such components to be similar in shape to the difference 
spectra derived in Fig. 4. 

%---------------------------------4.3Í---------------------------------------
\subsection{C3PO decomposition of the short-term variations}
%-------------------------------------------------------------------------------

%%%%%%%%%%%%%%%%%%figure7%%%%%%%%%%%%%%%%%%%
\begin{figure*}[t]
\epsscale{1}
\plotone{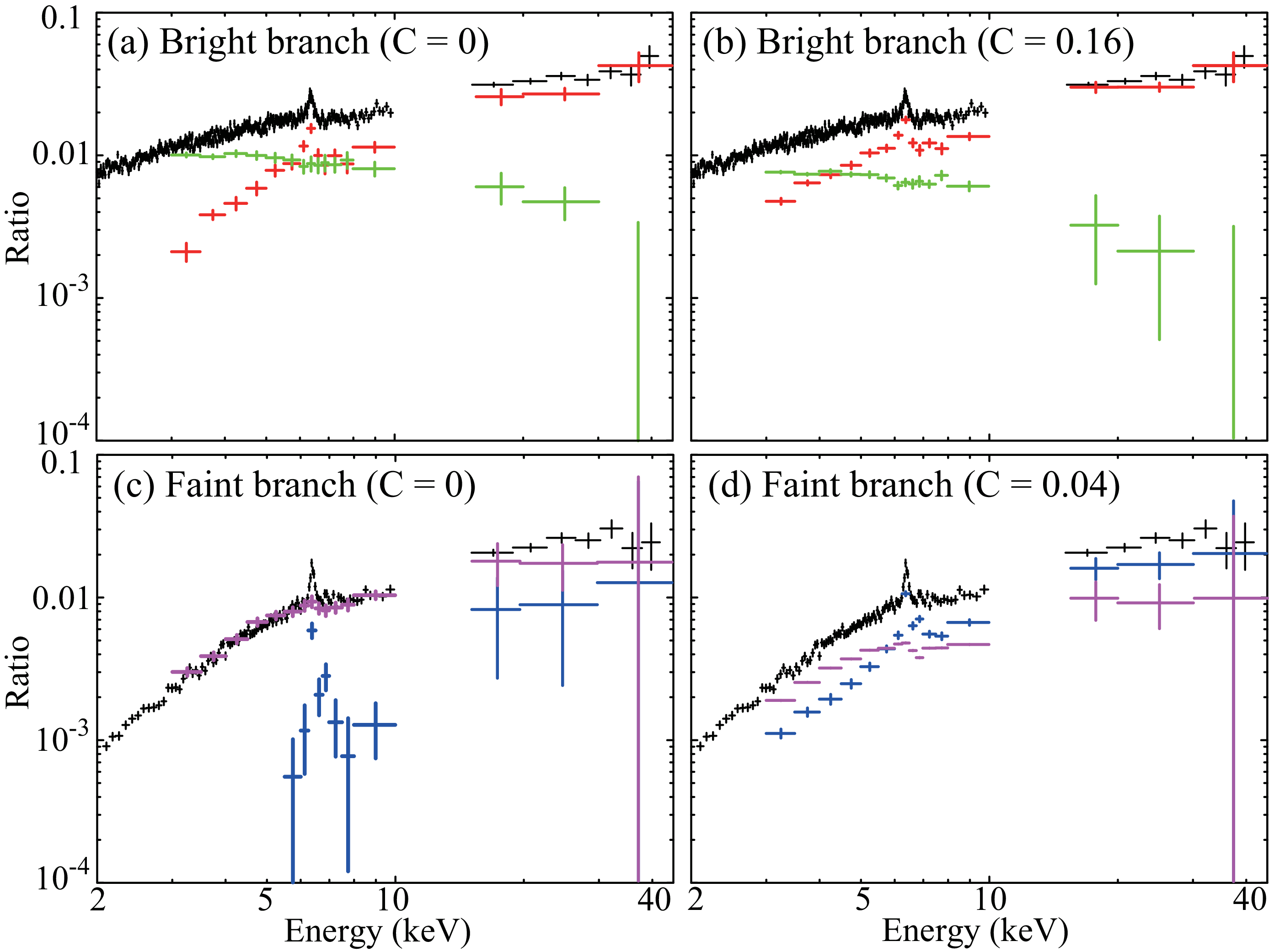}
\caption{Results of the C3PO application to the Bright- (panel a and b) 
and Faint-branch (panel c and d) data, consisting of the 1st+3rd and 
2nd+4th+5th+6th data sets, respectively. 
In panel (a) which assumes $C=0$, black shows the spectrum 
averaged over the 1st and 3rd observations, 
while green and red show the Bright-branch variable (BV) and stationary (BS) components, respectively.
Panel (b) is the same as panel (a), but assumes $C=0.16$. 
In panel (c) assuming $C=0$, black, purple, and blue show the averaged spectrum over the 
2nd, 4th, 5th, and 6th occasions, the Faint-branch variable (FV) and stationary (FS) ones, respectively. 
Panel (d) is the same as panel (c), but the case with $C=0.04$. 
 }
\end{figure*}
%%%%%%%%%%%%%%%%%%figure7%%%%%%%%%%%%%%%%%%%%

Having produced the CCP in Fig. 5, we applied the C3PO method developed 
by Noda et al. (2013b) to the entire assembly of the six data sets. 
After Noda et al. (2013b), the ``reference'' band was chosen to be 2--3 keV, where 
these objects generally exhibit the highest fractional variability. 
We then divided the 3--10 keV XIS and 
15--45 keV HXD-PIN ranges into 13 and 3 finer energy bands, respectively, with 
the energy boundaries at 3.0, 3.5, 4.0, 4.5, 5.0, 5.5, 6.0, 6.25, 6.5, 6.75, 7.0, 7.5, 8.0, and 10.0 in the XIS range, while 15.0, 20.0, 30.0, and 45.0 in the HXD-PIN range. 
As exemplified in Fig. 6, all the 16 finer-band CCPs exhibit the same break point at 
$x_{\rm B} \sim 0.16$ cnt s$^{-1}$ 
as that in Fig. 5. 
Hence, we can define the Bright- and Faint-branch in each finer-band CCP as well. 
Even if other energy bands are used as the reference, the derived CCPs generally exhibit 
this break point.  
%although in some cases the CCP slope above the break point is steeper than that below. 

The data in the two branches in each energy band were fitted separately 
with eq.(1), again utilizing the BCES algorithm. 
Results of the fits, summarized in Table 3, show that the short-term variation 
represented by the Bright branch has significantly 
positive values of $A$ and $B$ in all the CCPs, while the long-term one 
specified by the Faint branch has $B<0$ in lower energy bands. 

Following the recipe of the C3PO method given in Noda et al. (2013b), 
we decomposed the Bright-branch data (the 1st and 3rd data sets) into variable and 
stationary components, by utilizing the slopes $A$ and offsets $B$ in Table 3, respectively. 
Specifically, $y$ in each energy band  can be decomposed into a variable part, $Ax$, and  
a stationary part, $B$. 
By collecting the values of $Ax_{0}$ ($x_{0}$ being the average count rate 
in the reference band) and $B$, we can construct the variable-component 
spectrum and that of the stationary component, respectively, 
of which a sum exactly recovers the original time-averaged spectrum. 
Compared to the difference spectrum method (\S4.2) which can specify 
only the shape (but not normalization) of the variable component, 
the C3PO method allows us to determine both its shape and normalization. 

In extracting the short-term (Bright-branch) variable component, 
the slopes $A$ were multiplied by a factor $x_{\rm ob} =0.55$ cnt s$^{-1}$, 
which is the 2--3 keV count rate averaged over the two (the 1st and the 3rd) observations. 
As shown in Fig. 7(a) in green, this procedure gave a featureless spectrum with a rather steep slope. 
To compare it with the short-term difference spectrum shown in Fig. 4(a), 
we fitted it with the same \texttt{wabs * cutoffPL} model, where the column density of absorption, 
the photon index, and normalization were left free as before, 
while the cutoff energy was again fixed at 200 keV. 
The fit results in Table 2 confirm that the parameters of the variable component, except normalization, 
are all consistent with those of the short-term difference spectrum. 
As shown in Fig. 7(a) in red, 
the stationary component, in contrast,  was obtained
as a hard continuum, plus the prominent Fe-K$\alpha$ line which 
was originally in the time-averaged spectrum. 
Its fitting is carried out in \S4.4.

The above decomposition, utilizing eq.(1), implicitly assumes 
that the intensity $x$ in the reference (2--3 keV) band 
decreases to 0 as the source varies.
This assumption is not necessarily warranted.
Then, as already discussed in detail in Noda et al. (2013b, c),
we must  consider a non-zero ``intensity floor'' $C$,
to be identified with the minimum value of $x$ during the present observation,
and rewrite eq.(1) as
\begin{equation}
y = A(x-C)+B',
\end{equation}
with
\begin{equation}
B'=B+A C. 
\end{equation}
If we employ $C\sim 0.16$ cts s$^{-1}$ in the Bright branch (cf. Fig. 6),
the spectral decomposition results changes from Fig. 7(a) to (b).
Thus, the variable-component spectrum is affected
by a factor of $1 - C/x_0$, only in its normalization.
The stationary-component spectrum changes 
both in the shape and normalization.
However, the revised stationary spectrum is simply a sum
of that with $C=0$ and a fraction of the variable spectrum.
Thus, the two underlying components
remain unchanged even if $C \neq0$.
Below, we therefore employ the case with $C=0$ for simplicity.

%%%%%%%%%%%%%%%%%%figure8%%%%%%%%%%%%%%%%%%%
\begin{figure*}[t]
\epsscale{1}
\plotone{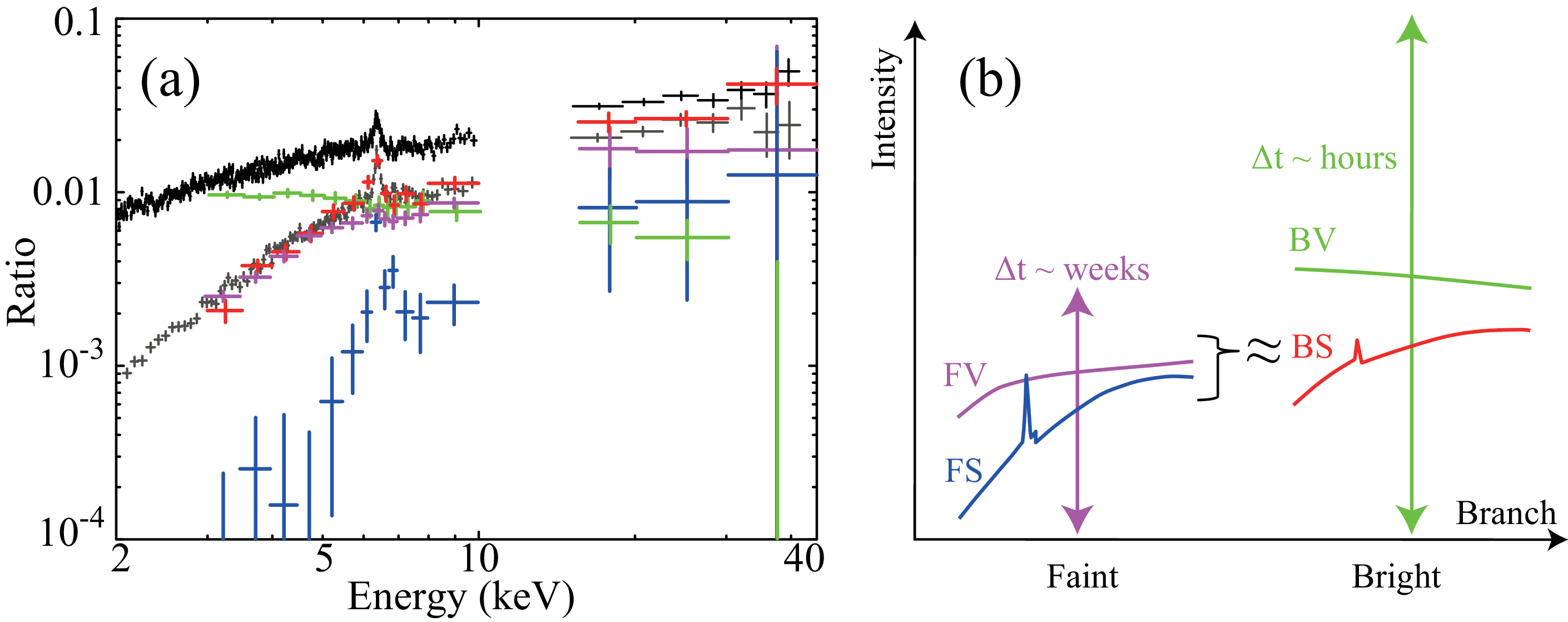}
\caption{(a) The same as a superposition of Fig. 7(a) and (b), but the FV (purple) 
and FS (blue) components were re-calculated using $C=0.01$ cnt s$^{-1}$. 
Grey shows the spectrum averaged over the Faint-branch observations  
(the same as black in Fig. 7b). 
(b) A schematic explanation of the relation (illustrated by solid curves) 
and variation (vertical solid arrows) of 
the four C3PO components, in the same colors as in panel (a). }
\end{figure*}
%%%%%%%%%%%%%%%%%%figure8%%%%%%%%%%%%%%%%%%%%

Thus, applying the C3PO method to   
the \textit{Suzaku} data acquired in the 1st and 3rd observations, the Bright-branch data
have been successfully decomposed into the variable and stationary parts, 
which we hereafter call BV (meaning Bright-variable; green in Fig. 7a) 
and BS (Bright-stationary; red in Fig.7a) components, respectively.
Except for normalization,
BV is essentially identical to the difference spectrum
derived from the short-term variation (Fig. 4a),
while BS contains reflection signals
(the Fe-K line and the hard X-ray hump).

%------------------------------------------------------------------------------
\subsection{C3PO studies of the long-term variations}
%------------------------------------------------------------------------------

The C3PO decomposition in \S 4.3 utilized the linear regression coefficients
($A$ and $B$ in Table 3) that are considered to represent the short-term variations,
namely, those obtained from the Bright branch in Fig. 5. 
Here, let us construct the C3PO spectra 
from the other set of coefficients in Table 3,
specified by the Faint-branch data distribution in the CCPs.
This enables us to study the long-term source changes among, as well as within,   
the fainter four data sets (the 2nd,  4th,  5th, and 6th).
In synthesizing the variable spectrum in reference to eq.(1),
we multiplied $A$ by a count rate of $x_{\rm of}=0.1$ cnt s$^{-1}$,
which is the reference 2--3 keV count rate 
averaged over the fainter 4 data sets.

Figure 7(c) shows the results of this C3PO analysis of the long-term variations.
Interestingly, the variable component (purple),
hereafter called Faint-branch variable (FV) component,
is very similar in spectral shape to the BS component (red in Fig. 7a),
except the lack of the prominent Fe-K$\alpha$ emission line. 
Instead, the Fe-K$\alpha$ line appears  in the Faint-branch stationary (FS) component,
shown in blue in Fig. 7(c).
The hard X-ray signals that were in the BS component (red in Fig. 7a)
were approximately halved by the C3PO procedure into FV and FS.
Although the latter (together with the Fe-K$\alpha$ line) must be the reflection hump,
the former is likely to be a part of the primary continuum. 

To confirm the suggested featureless property of FV,
we fitted it with the same model as used to study the long-term difference spectrum,
namely, \texttt{wabs * cutoffPL}, under the same fitting conditions.
As shown in Table 2, the fit has been successful,
and the results (except normalization) are fully consistent
with those from the difference spectrum for the long-term changes.
In other words, FV is considerably harder than BV, 
and is more absorbed. 

In Fig. 7(c), FS consists of  the prominent Fe-K$\alpha$ line and a hard X-ray hump, 
and is very similar in shape to the overall reflection from a non-relativistic material.
Actually, it can be fitted ($\chi^2/$d.o.f.=17.6/13) 
with a pure reflection model for an assumed input PL photon index of 1.7,
which yields an Fe abundance of $1.3^{+1.6}_{-0.5}$ Solar.
However, the spectrum declines too steeply below 6 keV, 
and lacks signals in the 3--5.5 keV band. 
This is clearly because the offset $B$ becomes negative therein (Table 3), 
which makes  FV and FS over- and under-estimated, respectively.

As already described in \S4.3, 
the C3PO decomposition has intrinsic uncertainty in the intensity flow $C$. 
In the Faint branch, $C$ can take a value   up to $0.04$ cnt s$^{-1}$  
which is the lowest count rate in the 2--3 keV band in the 4th observation. 
As $C$ is increased from $0$ to $0.04$ cnt s$^{-1}$, 
the Faint-branch decomposition changes  Fig. 7(c) to (d). 
As implied by eq. (3) and discussed by Noda et al. (2013b, 2013c), 
this stationary component for $C>0$ is nothing but a sum of the $C=0$ case 
and a fraction of FV. 
Therefore, the essence of the spectral decomposition does not depend on $C$. 
Since there is no a-priori way to specify this intensity floor, let us choose 
here $C=0.01$ cnt s$^{-1}$,  
which is the minimum to make all the offset values in the 3--45 keV band positive. 
[As clear from Fig. 8(a), the introduction of $C=0.01$ cnt s$^{-1}$ gives negligible 
effects on the Bright-branch C3PO analysis.] 
The corrected FV and FS components for the long-term variation are shown in Fig. 8(a). 
Thus, the positive intensity floor made FV and FS decreased and increased, respectively. 
As a result, 
FS (blue) has been fitted more successfully by the cold disk reflection model, 
with $\chi^2$/d.o.f.=12.7/13 and an Fe abundance of $1.0^{+0.5}_{-0.4}$ Solar. 
Below, we utilize these corrected FV and FS, derived with $C=0.01$ cnt s$^{-1}$, 
as the Faint-branch C3PO components.

%%%%%%%%%%%table 4%%%%%%%%%%%%
\renewcommand{\arraystretch}{1}
\begin{table*}[t]
 \caption{Parameters obtained in the triplet spectrum fits to the six NGC 3227 data sets.$^{\rm a}$ }
 \label{ngc3227_fit_parameter}
 \footnotesize
 \begin{center}
  \begin{tabular}{cccccccc}
  \hline\hline
  & & 1st & 2nd &  3rd & 4th & 5th & 6th \\\hline
  
  \multicolumn{3}{l}{Absorption to all}\\ [1.5ex]
  
  \texttt{wabs0} & $N_{\rm H}^{\rm b}$
   		& $1.6^{+0.3}_{-0.2}$
		&$1.6\pm0.5$
		&$3.2 \pm 0.4$
		&$1.4^{+0.4}_{-0.5}$
		&$<2.2$
		&$2.4^{+0.6}_{-0.8}$\\

 \texttt{zxipcf} & $N_{\rm H}^{\rm b}$
   		           & $23.2^{+27.9}_{-12.1}$
		            &--
		            &$3.0^{+23.9}_{-2.5}$
		            &--
		            &--
		            &--\\
		            
		             &log$\xi$ %(erg cm s$^{-1}$)
		             &$4.3^{+0.3}_{-0.2}$
		             &--
		             &$3.9^{+0.8}_{-0.6}$
		             &--
		             &--
		             &-- \\
		             
		             &Cvr frac.
		             &$1$ (fix)
		             &--
		             &$1$ (fix)
		             &--
		             &--
		             &--\\
		             
		             &$z$
		              &\multicolumn{6}{c}{0.0039 (fix)}\\\hline

 \multicolumn{3}{l}{BV}\\ [1.5ex]
  \texttt{cutoffPL0}   & $\Gamma_0$%
                       & $2.34 \pm 0.09$
                       & --
                       &$2.49^{+0.12}_{-0.11}$
                       &--
                       &--
                       &--\\

		       &$E_{\rm cut}$~(keV)
		       &\multicolumn{6}{c}{200 (fix)}\\

                    & $N_\mathrm{PL0}^{\rm c}$%sa
                       & $1.84^{+0.26}_{-0.24}$
                       &--
                       &$1.05^{+0.23}_{-0.18}$
                       &--
                       &-- 
                       &--\\\hline
                       
  \multicolumn{3}{l}{Reflection component (FS)}\\ [1.5ex]                      
                       
   \texttt{pexmon} & $\Gamma_{\rm ref}$%
                          &$=\Gamma_{\rm PL}$
                          &$2$ (fix)
                          &$=\Gamma_{\rm PL}$
                          &$2$ (fix)
                          &$2$ (fix)
                          &$2$ (fix) \\

		       &$E_{\rm cut}$~(keV)
		       &\multicolumn{6}{c}{200 (fix)}\\
   
   			& $f_\mathrm{ref}$%
                          &$0.9 \pm 0.2$
                          &$1$ (fix)
                          &$1.1 \pm 0.1$
                          &$1$ (fix)
                          &$1$ (fix)
                          &$1$ (fix)\\
 
 		       &$z$
		       &\multicolumn{6}{c}{0.0039 (fix)}\\

 		       &$A$~($Z_{\odot}$)
		       &\multicolumn{6}{c}{1 (fix)}\\

 		       &$A_{\rm Fe}$~($Z_{\rm{Fe}, \odot}$)
		       &\multicolumn{6}{c}{1 (fix)}\\

 		       &$i$~(degree)
		       &\multicolumn{6}{c}{60 (fix)}\\

                          & $N_\mathrm{ref}^{\rm d}$%
                          &$=N_{\rm PL0}$
                          &$1.50 \pm 0.13$
                          &$=N_{\rm PL0}$
                          &$1.32 \pm 0.09$
                          &$1.38\pm 0.11$
                          &$1.22 \pm 0.12$\\\hline
                          
  \multicolumn{3}{l}{FV}\\ [1.5ex]   
 			
  \texttt{pcfabs1} & $N_{\rm H}^{\rm b}$
   		& $9.7^{+2.9}_{-1.8}$
		&$21.1^{+4.2}_{-3.6}$
		&$11.4^{+1.1}_{-1.0}$
		&$40.0^{+7.4}_{-7.0}$
		&$9.8^{+1.3}_{-0.8}$
		&$15.3^{+2.9}_{-2.4}$\\			
		
		  &  Cvr frac.
		   &$>0.64$
		&$0.63^{+0.08}_{-0.03}$
		&$>0.57$
		&$0.65 \pm 0.05$
		&$0.80 \pm 0.03$
		&$0.69^{+0.02}_{-0.04}$\\

  \texttt{cutoffPL1}   & $\Gamma_1$%
                       & $1.48^{+0.04}_{-0.05}$
                       & $1.41^{+0.07}_{-0.06}$
                       &$1.66^{+0.05}_{-0.04}$
                       &$1.31\pm0.12$
                       &$1.67 \pm 0.06$
                       &$1.66 \pm 0.08$\\

		       &$E_{\rm cut}$~(keV)
		       &\multicolumn{6}{c}{200 (fix)}\\

                    & $N_\mathrm{PL1}^{\rm e}$%
                       & $0.42 \pm 0.06$
                       & $0.26^{+0.06}_{-0.04}$
                       &$0.63 \pm 0.06$
                       &$0.12\pm0.01$
                       &$0.56\pm 0.07$ 
                       &$0.38^{+0.06}_{-0.05}$\\ \hline

   $\chi^{2}$/d.o.f.& & 419.45/355  & 180.70/155 & 252.35/221 & 132.25/109& 243.21/249 & 142.04/136\\
      \hline\hline
      
  \end{tabular}
 \end{center}
   
  	{\small
	$^{\rm a}$ In the 4th triplet fit, FV was changed into FV\_4th.  \\
	$^{\rm b}$ Equivalent hydrogen column density in  $10^{22}$ cm$^{-2}$. \\
         $^{\rm c}$ The \texttt{cutoffPL0} normalization at 1 keV, in units of $10^{-2}$~photons~keV$^{-1}$~cm$^{-2}$~s$^{-1}$~at 1 keV.\\
         $^{\rm d}$ The \texttt{pexmon} normalization at 1 keV, in units of $10^{-2}$~photons~keV$^{-1}$~cm$^{-2}$~s$^{-1}$~at 1 keV.\\
      $^{\rm e}$ The \texttt{cutoffPL1} normalization at 1 keV, in units of $10^{-2}$~photons~keV$^{-1}$~cm$^{-2}$~s$^{-1}$~at 1 keV.}
   
\end{table*}
\renewcommand{\arraystretch}{1}
%%%%%%%%%%%table 4%%%%%%%%%%%%

%%%%%%%%%%%%%%%%%%figure9%%%%%%%%%%%%%%%%%%%
\begin{figure*}[p]
\epsscale{1}
\plotone{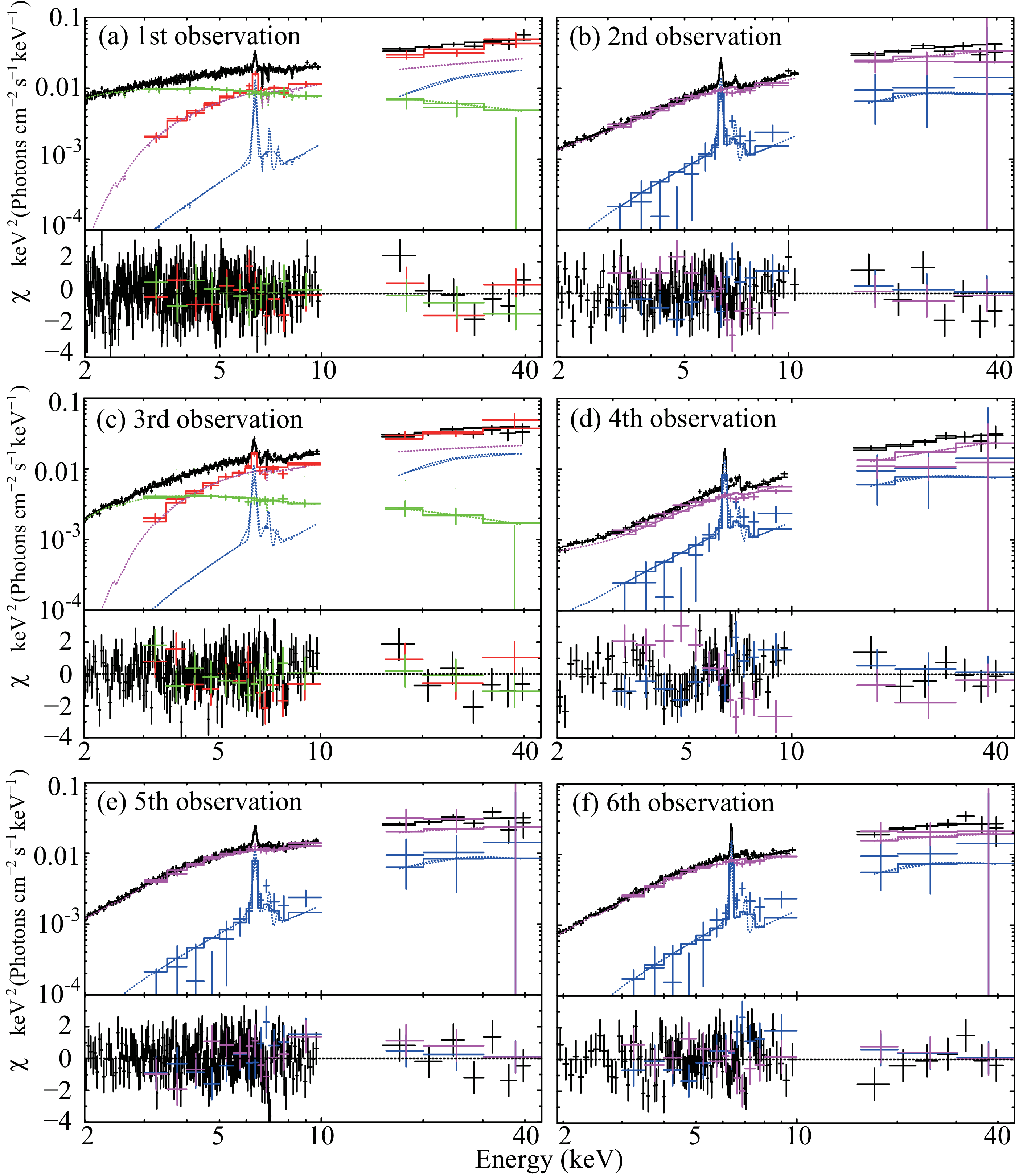}
\caption{Simultaneous fits to the time-averaged (black), BV (green), BS (red), 
FV (purple), and FS (blue) spectra. The 1st (panel a) and 3rd (panel c) data sets are fitted with 
\texttt{model\_BV} and \texttt{model\_BS} defined in the text,  
while the other spectra with \texttt{model\_FV} and \texttt{model\_FS}. }
\end{figure*}
%%%%%%%%%%%%%%%%%%figure9%%%%%%%%%%%%%%%%%%%%

In Fig. 8(a), the Bright-branch spectra (the same as Fig. 7a; BV in green, BS in red, and 
the total in black) are superposed for comparison. 
We reconfirm 
the close resemblance of BS (red) and FV (purple) except the Fe-K$\alpha$ line 
(present in BS but absent in FV)
and some differences in the hard X-ray hump. 
There, the time-averaged spectrum in the Faint branch (black in Fig. 7c) is also reproduced 
in grey. 
Surprisingly, BS is nearly identical, both in spectral shape and normalization, 
to the Faint-branch time-averaged spectrum, including the Fe-K$\alpha$ emission line and 
the Compton hump. 
From these characteristics, 
we arrive at a new interpretation of the spectral components of NGC 3227, 
which is illustrated in Fig. 8(b) and summarized below. 
\begin{enumerate}
\item In the Bright branch, BV dominates, 
and varies on short time scales ($\sim$several hours) without significant spectral shape changes. 
\item When BV disappears, BS which is equivalent to FV+FS starts dominating, 
with its intensity kept approximately constant on short time scales.
The source then changes into the Faint branch. 
\item In the Faint-branch state, FV varies on a longer time scale ($\gtrsim1$ week), 
while FS remains unchanged for $\gtrsim1$ month. 
\end{enumerate}

%---------------------------------------------------------------------------
\section{Overall spectral analyses}
%--------------------------------------------------------------------------

So far, the C3PO analysis has been conducted in two steps.
First, the time-averaged spectra were decomposed 
into their variable and stationary parts (\S 4.3, \S 4.4).
Then, (some of) the derived C3PO components were examined through 
standard spectral model fitting procedures (also in \S 4.3, \S 4.4).
In the present section, we proceed to the third step, 
to be called ``triplet spectrum fitting" (Noda et al. 2013a, 2013b).
That is, we \emph{simultaneously} fit the variable component, 
the stationary component, and the time-averaged spectrum,
by an appropriate spectral model, by another model,
and by the sum of the two models, respectively.
This is because the variable and stationary parts,
when summed together, 
should recover, by definition,  the time-averaged spectrum.
By thus incorporating the C3PO results,
we can significantly reduce the spectral modeling ambiguity
associated with the overall model composition.

This triplet fitting procedure is also much more constraining  
than separately quantifying the variable and stationary spectra individually, 
because the time-averaged spectrum retains high statistics
which are partially lost in the  two C3PO spectra.
To be rigorous, the errors in $A$ and  $B$ are anti-correlated,
and hence the two C3PO spectra are not completely independent.
Although the degrees of freedom in the simultaneous fit 
should be reduced accordingly,
we neglect this effect;
this is because the combined degree of freedom is dominated
by that of the time-averaged spectrum (with much larger bin numbers).
In addition, this simplification makes our study more conservative,
in the sense that it tends to underestimate differences 
in the fit goodness among different model compositions.

In short, the triplet spectral fitting can attain much higher statistical accuracy than 
analyzing only the two C3PO-derived spectra. It can also significantly reduce systematic 
modeling ambiguity compared to the more conventional analysis using the 
time-averaged spectrum alone. 

%%%%%%%%%%%%%%%%%%figure10%%%%%%%%%%%%%%%%%%%
\begin{figure}[t]
\epsscale{1.1}
\plotone{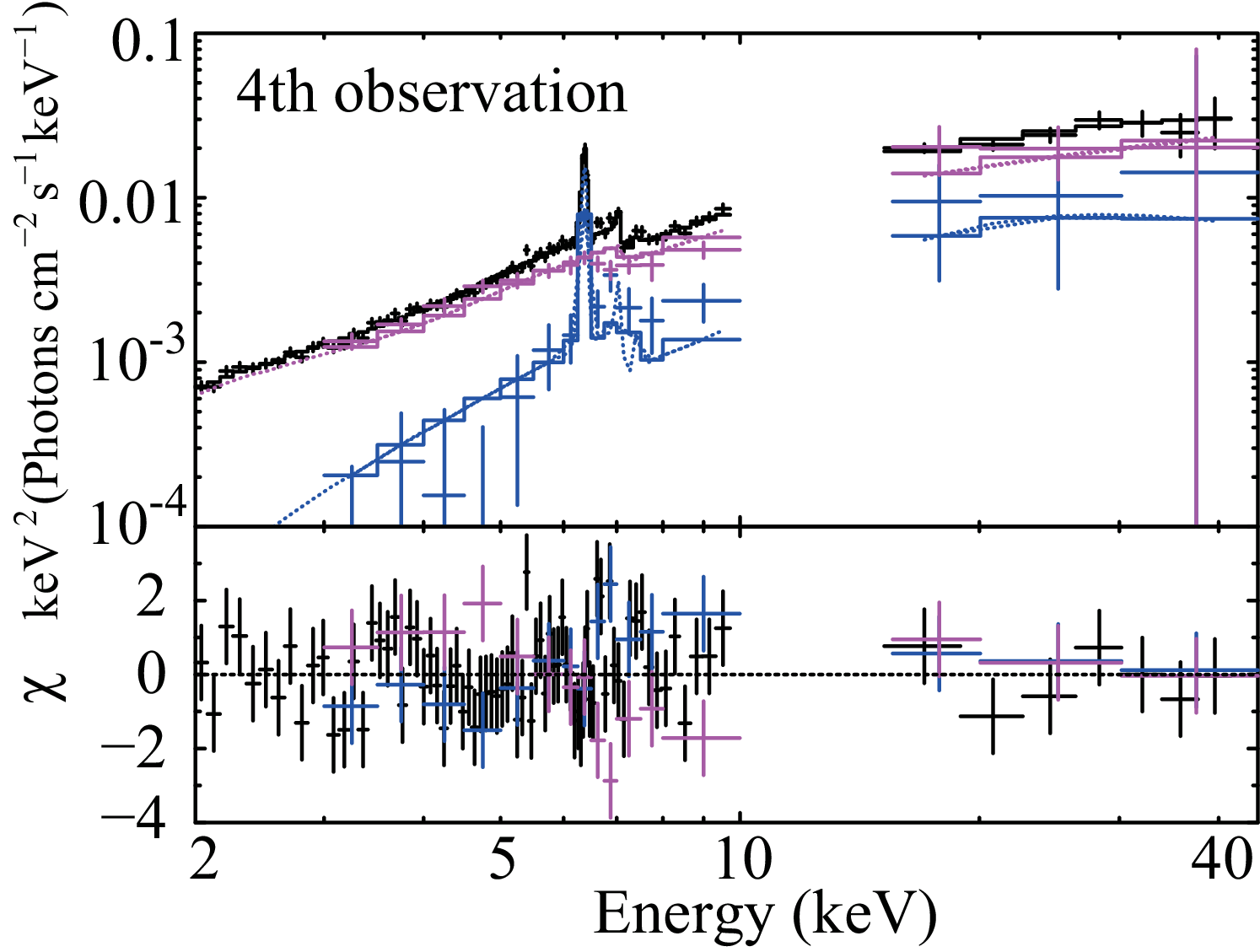}
\caption{The same 4th spectrum as Fig. 9(d), but the FV component 
was changed into FV\_4th (see text). }
\end{figure}
%%%%%%%%%%%%%%%%%%figure10%%%%%%%%%%%%%%%%%%%%

Let us apply the triplet spectrum fitting to the 1st and 3rd (Bright branch) data sets, 
following the same analysis as conducted on NGC 3516  by Noda et al. (2013b). 
Preparing \texttt{model\_BV $\equiv$ wabs0 * zxipcf * cutoffPL0} and 
\texttt{model\_BS $\equiv$ wabs0 * (pexmon + wabs1 * cutoffPL1)}, 
we fitted BV, BS, and the time-averaged spectrum, 
with \texttt{model\_BV}, \texttt{model\_BS}, and 
\texttt{model\_BV + model\_BS}, respectively. 
In the fits, the column densities of \texttt{wabs0} and \texttt{wabs1} were left free. 
In \texttt{cutoffPL0} and \texttt{cutoffPL1}, 
which represent the main continua in the two  C3PO spectra, 
the photon indices and the normalizations were left free, 
while their cutoff energies  were both fixed at 200 keV. 
In the \texttt{pexmon} model, 
the photon index of the incident PL, cutoff energy, 
and the normalization were tied to those of \texttt{cutoffPL0}, assuming 
a condition wherein the dominant PL, \texttt{cutoffPL0}, produce the reflection component 
represented by \texttt{pexmon} (item 2 in \S4.4);   
the abundance of Fe and lighter elements, inclination, and the redshift were fixed at 
1 Solar, 60$^{\circ}$, and 0.00386, respectively, 
while the reflection fraction relative to \texttt{cutoffPL0} was left free.  
The \texttt{zxipcf} model represents partial covering absorption by partially ionized materials 
(Reeves et al. 2008),  
wherein the column density and the ionization parameter were left free, 
while the covering fraction and redshift were fixed at 1 and 0.0039, respectively. 
The introduction of this factor is not for a partial absorption effect, 
but to allow possible presence of warm absorbers as often found in Seyfert galaxies.

As shown in Fig. 9(a) and Fig. 9(c), the triplet fits were successful both in the 1st and 3rd data sets, 
giving parameters summarized in Table 4. 
The photon index of BV became $\Gamma_0 \sim 2.3$, while that of the additional PL, 
which is part of BS and later becomes equivalent to FV, turned out to be significantly 
harder with $\Gamma_1 \sim 1.5$. 
In addition, the absorption working on the additional PL became significantly higher 
($N_{\rm H1} \sim 10^{23}$ cm$^{-2}$) than that on BV. 
The change in the time-averaged spectrum from the 1st to the 3rd data sets can be 
explained solely by an intensity decrease of BV, with little changes (within errors) 
in its shape.

Next, we applied the same triplet spectrum fitting  
individually to the 2nd, 4th, 5th, and 6th data sets (Faint branch). 
We employed \texttt{model\_FV $\equiv$ wabs0 * wabs1 * cutoffPL1} and 
\texttt{model\_FS $\equiv$ wabs0 * pexmon}, 
following the successful fit to FV in \S4.4
and the expectation of FS to be a pure neutral reflection component. 
Then, FV, FS, and the time-averaged spectrum 
of each observation were fitted simultaneously with \texttt{model\_FV}, \texttt{model\_FS}, 
and \texttt{model\_FV + model\_FS}, respectively. 
The fits were conducted under the same parameter conditions 
as those in the fits to the 1st and 3rd data sets, 
except that  
the reflection fraction and normalization of \texttt{pexmon} 
were fixed at 1 and left free, respectively. 
The photon index of the incident PL of \texttt{pexmon} was fixed at 2, 
assuming the reflection component to be generated by illumination of 
both FV ($\Gamma_1 \sim 1.6$) and afterglow of BV ($\Gamma_0 \sim 2.3$).

Actually, 
the fits all failed with $\chi^2/$d.o.f. = 432.5/156, 330.9/110, 433.3/250, and 321.1/138
in the 2nd, 4th, 5th, and 6th data sets, respectively, 
all due to positive residuals below $\sim 3$ keV. 
Even though the C3PO-derived variable and stationary components 
can be explained individually as shown in \S4.4, 
the joint triplet fits were unsuccessful 
mainly because  the time-averaged spectra have less convex (and even
concave in the 4th) shapes than is represented by the absorbed PL model. 
Then, we changed the model describing the FV component  
into \texttt{model\_FV$_{\rm P}$ $\equiv$ wabs0 * (pcfabs * cutoffPL1)}, where \texttt{pcfabs} represents 
a partially covering absorption with the column density and covering factor left free. 
As shown in Fig. 9 and Table 4, the fits to the Faint-branch data sets, 
except for the 4th observation with $\chi^2/$d.o.f=184.35/108, has become acceptable, 
giving a covering fraction somewhat larger than 
$50$\%. 
Following the results of the Faint-branch datasets, 
we replaced \texttt{wabs1} in \texttt{model\_BS} by \texttt{pcfabs1} 
with the covering fraction left free, 
and changed the model for the BS spectra 
into \texttt{model\_BS$_{\rm P}$ $\equiv$ wabs0 * (pexmon + pcfabs1 * cutoffPL1)}. 
As a result, the covering fraction has become consistent with 1, giving lower limits of $\sim0.6$, 
in both triplet fits to the 1st and 3rd datasets as shown in Table 4.
The fits hence gave almost same parameter values 
as those obtained by the fits with \texttt{model\_BS}.

Let us further examine the 4th observation, when the source was faintest. 
As suggested in Fig. 9(d) by residuals in purple, FV in this particular observations 
may have a somewhat different shape from those of the others. 
Actually, the data points of this particular observation 
are slightly deviated in Fig. 5 and Fig. 6 from the Faint-branch straight line 
described by eq. (1), in which $A$ and $B$ are determined jointly by the four data sets. 
In other words, these parameters, particularly $A$, need to be re-determined. 
We hence applied the C3PO method to the 4th data set alone (Fig. 6 purple), 
with the offsets $B$ fixed at the values shown in the ``Long-term'' column in Table 3. 
The intensity floor was again assumed to be $C=0.01$ (\S4.4). 
The results are presented in Fig. 10, 
where the re-calculated FV (renamed FV\_4th) spectrum is shown in purple. 
The triplet spectral fitting to the 4th data set thus became acceptable with $\chi^2/$d.o.f.=132.25/109, 
and the obtained parameters are shown in Table 4. 
The column density of the partial absorption of the 4th observation was found 
to be significantly larger than in the other observations, changing FV into FV\_4th. 
To summarize, the difference among the 2nd, 5th and 6th time-averaged spectra can be attributed 
mostly to intensity changes of FV alone, while a slight spectral hardening occurred on  
the 4th occasion.

%%%%%%%%%%%table 5%%%%%%%%%%%%
\renewcommand{\arraystretch}{1}
\begin{table}[t]
 \caption{Same as Table \ref{ngc3227_fit_parameter}, but  the 2nd spectra are fitted 
         with \texttt{model\_FV' = wabs * kdblur * reflionx} and  \texttt{model\_FV" = 
	wabs * (cutoffPL1 + comptt)}.}
 \label{ngc3227_2nd_compton_fit_table}
 \small
 \begin{center}
  \begin{tabular}{ccc}
   \hline\hline
  & & 2nd \\\hline

  \texttt{wabs} & $N_{\rm H}^{\rm a}$
   		& $3.1$\\

   \texttt{pexmon} & $\Gamma_{\rm ref}$%
                          &$2$ (fix)\\
                         
                          & $E_{\rm cut}$~(keV)%
                          &$200$ (fix)\\
                          
                          & $f_{\rm ref}$%
                          &$1$ (fix)\\

                          & $z$%
                          &$0.0039$ (fix)\\

                          & $A$~($Z_{\odot}$)%
                          &$1$ (fix)\\

                          & $A_{\rm Fe}$~($Z_{\rm{Fe}, \odot}$)%
                          &$1$ (fix)\\

                          & $i$~(degree)%
                          &$60$ (fix)\\
                          
                          & $N_\mathrm{ref}^{\rm b}$%
                          &$1.6$\\[1.5ex]

  \texttt{kdblur} &   $q$
		   			& $5.4$\\
					
				 & $R_{\rm in}$ ($R_{\rm g}$)
		   		&$1.4$\\
				
			 & $R_{\rm out}$ ($R_{\rm g}$)
		   		&$400$\\					
  
  				 & $i$ (degree)
		   		&$60$ (fix)\\		
  
 \texttt{reflionx} & $A_{\rm Fe}$ ($Z_{\odot}$)
		   &$1$ (fix)\\		
  
  			&$\Gamma$ (keV)
   		& $2.3$ (fix)\\			

  			&$\xi$ (erg cm s$^{-1}$)
   		& $25.5$ \\

		   & $N_{\rm reflionx}^{\rm c}$
		   &$6.8$\\[1.5ex]

   $\chi^{2}$/d.o.f.& & 299.2/156 \\\hline\hline

  \texttt{wabs} & $N_{\rm H}^{\rm a}$
   		& $0.17$\\

   \texttt{pexmon} & $\Gamma_{\rm ref}$%
                          &$2$ (fix)\\
                         
                          & $E_{\rm cut}$~(keV)%
                          &$200$ (fix)\\
                          
                          & $f_{\rm ref}$%
                          &$1$ (fix)\\

                          & $z$%
                          &$0.0039$ (fix)\\

                          & $A$~($Z_{\odot}$)%
                          &$1$ (fix)\\

                          & $A_{\rm Fe}$~($Z_{\rm{Fe}, \odot}$)%
                          &$1$ (fix)\\

                          & $i$~(degree)%
                          &$60$ (fix)\\
                          & $N_\mathrm{ref}^{\rm b}$%
                          &$1.6$\\

  \texttt{cutoffPL1} 
  
                            & $\Gamma_1$%
                          &$1.50$ \\
  
  		& $E_{\rm cut}$ (keV)
   		& $200$ (fix)\\				
		   
		   & $N_{\rm PL1}^{\rm d}$
		   &$0.48$\\[1.5ex]
                                           
    \texttt{comptt} 
                            & $z$%
                          &$0.0039$ (fix)\\
  
  		& $T_0$ (keV)
   		& $1.67$ \\			
		
		  & $T_{\rm e}$ (keV)
		   &$200$ (fix)\\			

		  & $\tau$
		   &$0.62$\\		
		   
		   & $N_{\rm Comp}^{\rm e}$
		   &$0.27$\\[1.5ex]

   $\chi^{2}$/d.o.f.& & 244.92/155 \\
      \hline\hline
      
  \end{tabular}
 \end{center}
   
  	{\small
	$^{\rm a}$ Equivalent hydrogen column density in  $10^{22}$ cm$^{-2}$. \\
         $^{\rm b}$ The \texttt{pexmon} normalization at 1 keV, in units of $10^{-2}$~photons~keV$^{-1}$~cm$^{-2}$~s$^{-1}$~at 1 keV.\\
      $^{\rm c}$ The \texttt{reflionx} normalization in units of $10^{-6}$.\\
   $^{\rm d}$ The \texttt{cutoffPL1} normalization at 1 keV, in units of $10^{-2}$~photons~keV$^{-1}$~cm$^{-2}$~s$^{-1}$~at 1 keV.\\
$^{\rm e}$ The \texttt{comptt} normalization, in units of $10^{-5}$
         photons~keV$^{-1}$~cm$^{-2}$~s$^{-1}$~at 1 keV.}

\end{table}
\renewcommand{\arraystretch}{1}
%%%%%%%%%%%table 5%%%%%%%%%%%%

%----------------------------------------------------------------------------------------
\section{Examination of alternative interpretations}
%-----------------------------------------------------------------------------------------

Although we have so far been modeling FV with a partially-absorbed PL, 
other interpretations still remain possible. 
Because NGC 3227 has periods when the rapidly variable PL  (i.e., BV) disappears, 
we can utilize such Faint-branch data to test other possible models 
under the least contamination by BV. 
One popular model to explain such a concave spectrum 
is a relativistically-blurred and ionized reflection interpretation,  
which invokes an extremely spinning Kerr BH
and a strong ``light bending'' effect (e.g., Miniutti \& Fabian 2004; Miniutti et al. 2007). 
To test this interpretation, 
we replaced the model for FS  by \texttt{model\_FV' = wabs * kdblur * reflionx}, 
where \texttt{kdblur} and \texttt{reflionx} represent 
a convolution model for relativistic effects (Laor et al. 1991)
and a model for a reflection component generated at 
an ionized relativistic accretion disk (Ross \& Fabian 2005), respectively. 
In this model, the outer disk radius was fixed at 400 $R_{\rm g}$, 
the photon index of the incident PL at 2.0, the lighter element abundance at 1 Solar, 
and the inclination at  60$^{\circ}$, while the other parameters were left free. 
The fit was however unsuccessful as shown in Fig. 11(a) and Table 5, 
with $\chi^2/$d.o.f.=299.2/156 (cf. 180.7/155 previously), 
mainly due to large positive residuals produced 
by the time-averaged spectrum in the 7--10 keV band.  
Thus, the relativistically-blurred reflection interpretation 
fails to explain the FV component of NGC 3227. 

Can the soft energy drop seen in FV be explained without the absorption/reflection effects?
A thermal Comptonization with a high seed photon temperature
(e.g., $\sim2$ keV)  becomes a candidate, 
because  the spectrum is strongly enhanced by Comptonization
at higher-energy side of such an initial seed-photon distribution,
while its  lower-energy side must remain rather unchanged
and will be directly visible in our spectrum.
Since the original seed-photon distribution in this Rayleight-Jeans regime
is much harder than the PLs we are considering ($\Gamma \sim $1.4--1.6),
the emergent Comptonized spectrum will naturally exhibit a low-energy roll over.
When some of seed photons of the thermal Comptonization are generated 
as soft X-rays in higher temperature regions somewhere in the system, 
while the rest, as usual, by a standard accretion disk, 
a concavely-shaped Compton continuum may appear, as 
a sum of the two PL components with and without the soft X-ray cutoff. 
To examine this probability, 
we replaced the model for FV by \texttt{model\_FV" = wabs * (cutoffPL1 + comptt)}, 
where \texttt{comptt}, a thermal Comptonization continuum, 
is alternative to the absorbed PL in \texttt{model\_FV}. 
The cutoff energy in \texttt{cutoffPL1} was fixed again at 200 keV, 
while the photon index and the normalization were left free. 
The electron temperature and the redshift of \texttt{comptt} 
were fixed at 200 keV and 0.00386, respectively, 
while its seed photon temperature, optical depth and the normalization were left free. 
As shown in Fig. 11(b) and Table 5, 
the low-energy drop of FV was successfully reproduced 
in the way without using absorption by a 
relatively high seed-photon temperature (1.67 keV), but 
the fit itself was unsuccessful with $\chi^2/$d.o.f.=244.92/155, mainly because of 
negative residuals around $\sim7.1$ keV. 
This  structure is considered to correspond to a neutral Fe-K absorption edge 
produced by a neutral absorber, and the failure of this high-temperature seed Compton  
model is caused by its inability to explain the Fe-K edge seen in the actual data. 
In other words, the FV component is best reproduced by a strongly (and partially) 
absorbed PL, and its low energy cutoff is in fact due to photoelectric absorption 
by nearly neutral matter. 

%%%%%%%%%%%%%%%%%%figure11%%%%%%%%%%%%%%%%%%%
\begin{figure*} [t]
\epsscale{1}
\plotone{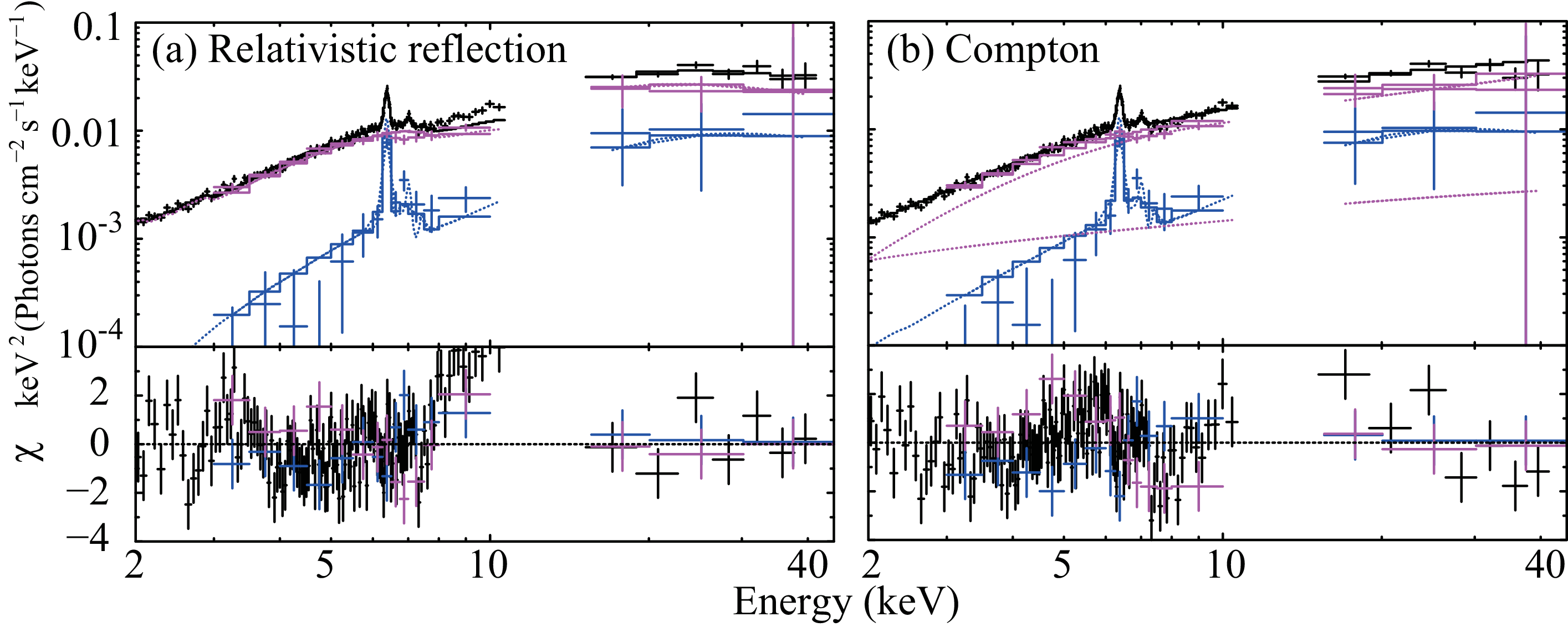}
\caption{Same as Figure 9(b), but the model for the FV component
is replaced with \texttt{model\_FV' = wabs * kdblur * reflionx} (panel a) and 
  \texttt{model\_FV" = wabs * (cutoffPL1+ comptt)} (panel b).   }
\end{figure*}
%%%%%%%%%%%%%%%%%%figure11%%%%%%%%%%%%%%%%%%%%

%=============================
\section{Discussion and Conclusion}
%=============================

\subsection{Summary of the results}

We analyzed the six data sets of NGC 3227 obtained by \textit{Suzaku} in 2008, 
and applied the two methods of variability-assisted spectroscopy; 
the difference spectrum analysis (\S4.1) and 
the C3PO method (\S4.2--\S4.4; Noda et al. 2013b).  
As a results, the source behavior was found to differ significantly above (the Bright branch) 
and below (the Faint branch) the threshold count rate of 
$x_{\rm B} \sim0.15$ cnt s$^{-1}$ in 2--3 keV ($\sim 0.8$ cnt s$^{-1}$ in 3--10 keV).  
These rates translate to a 2--10 keV threshold luminosity 
of $L_{\rm th} \sim6.6\times 10^{41}$ erg s$^{-1}$. 
Furthermore, we identified four spectral components; BV, BS, FV, and FS. 
Since BS $\approx$ FV+FS (Fig. 8b), the entire 2--45 keV emission is considered to consist 
of the following three components. 

\begin{enumerate}
\item A rapidly variable steep PL-like continuum, identified as BV in \S4.1, \S4.3 and \S5. 
It suffers relatively low absorption.

\item A slowly variable hard PL component, found as FV (a part of BS) in \S4.1 \S4.4, and \S5.  
Its spectral shape is strongly affected by neutral absorption.

\item A reflection component accompanied by a narrow Fe-K$\alpha$ emission 
line, identified as FS (Fig. 8b).  
It is emitted by cold and neutral materials without strong relativistic effects.

\end{enumerate}

%-------------------------------------------------------
\subsection{The Bright-branch Variable Emission}
%-------------------------------------------------------

The BV component (component 1 in \S7.1) 
has a relatively steep spectral slope with a photon index of $\sim 2.3$, 
and is weakly absorbed with a column density of $\lesssim 10^{22}$ cm$^{-2}$. 
It appears only when the source is in the Bright branch with $L \gtrsim L_{\rm th}$ 
(with $L$ being 2--10 keV luminosity), and the source brightening above $L_{\rm th}$ 
is almost solely attributed to the increase of this component. 
Its spectral shape does not change when it varies in intensity 
by almost an order magnitude, 
making the CCP distributions linear in the Bright branch. 
It is highly variable on a short timescale from several hundreds ksec down to $\sim 5$ ksec, 
namely $\sim 25~R_{\rm g}/c$.  

Past X-ray studies reported repeatedly 
(e.g., Markowitz \& Edelson 2001; Caballero-Garcia et al. 2012; Soldi et al. 2013) that 
Seyfert galaxies exhibit spectral steepening when they become brighter, or ``softer when brighter''. 
In our view, this can be explained naturally by an increasing contribution of this BV continuum, 
while the other two components (2 and 3 in \S7.1) with harder spectra remains 
approximately constant. 

The spectral shape of BV is relatively similar to the PL-like continuum with $\Gamma \sim 2.5$ 
(the hard tail) of BH binaries found in their high/soft state  
(e.g, Gierli{\'n}ski et al. 1999; Remillard \& McClintock 2006), 
and the entire ($\gtrsim2$ keV) X-ray emission from narrow-line Seyfert I galaxies 
with $\Gamma \lesssim 2.3$ (e.g., Laor et al. 1994; Boller et al. 1996). 
Furthermore, these steep-PL components all appear when the source has a relatively 
high luminosity. 
Thus, the BV component may represent the Seyfert emission at relatively high accretion rates.  
In fact, the type I Seyfert NGC 3516 exhibited a very similar component with $\Gamma \sim 2.2$, 
variable and weakly absorbed, when it was in a bright state in 2005 (Noda et al. 2013b). 
At such a high accretion rate, 
a standard accretion disk is expected to extend down to inner regions close to the BH, 
and BV may originate therein, clearly as a primary X-ray emission:  
discussion continues in \S7.5.

%-------------------------------------------------------
\subsection{The Faint-branch Variable Continuum}
%-------------------------------------------------------

The FV component persisted through the entire 6 observations. 
It was identified as the stationary component when $L>L_{\rm th}$ 
and hence  BV is strong (the 1st and 3rd data), while it carried 
a major fraction of the 2--45 keV emission at $L<L_{\rm th}$. 
Compared to BV, 
it exhibits a significantly harder spectral shape with a photon index of $\sim1.6$, 
and is affected by a relatively strong neutral absorption 
with a column density of $\sim10^{23}$ cm$^{-2}$. 
While this component is much less variable than BV, 
it varied significantly by a factor of two on a long time scale of $\gtrsim1$ week, 
and produced the almost linear CCPs in the Faint branch. 
Unlike the reflection component (component 3 in \S7.1), 
it lacks the Fe-K$\alpha$ emission line, 
and can be reproduced by neither a neutral nor relativistically-smeared reflection.  
It cannot be a Comptonization with a high seed photon temperature, either. 
More generally, this component cannot be secondary signals produced by BV, 
since it varied quite independently of BV that makes up a major 
part of the primary emission in the Bright branch. 
In particular, it persisted even when BV disappeared. 
We hence regard it as a new primary emission from the central engine. 

In previous studies of type I Seyferts, this component was 
interpreted presumably as a partially absorbed part of the dominant BV continuum. 
However, its variability timescale is rather different from that of BV, 
in contradiction to what the strongly absorbed part of BV should have. 
Furthermore, FV would then have $\Gamma \sim 2.3$ as well, in disagreement with our observation. 
We thus reconfirm our inference made above, 
that the central engine of NGC 3227 (and probably of 
similar objects) is emitting FV as another primary continuum 
that is independent of the BV component. 
The absorbed hard PL component of NGC 3516, 
obtained in the 2005 \textit{Suzaku} data set  by Noda et al. (2013b), 
can be identified with FV in NGC 3227, because of its resemblances in 
the spectral slope and the amount of absorption. 
Therefore, this $\Gamma \sim 1.6$ component is expected to be common among 
type I Seyferts when they are relatively dim.

The spectral characteristics of FV may be similar to those of the thermal Comptonization continuum 
of BH binaries in the low/hard state (e.g., Makishima et al. 2008; Yamada et al. 2013) 
or low-luminosity AGNs (e.g., Terashima et al. 2002; 
Terashima \& Wilson 2003), of which the photon index is $\sim 1.7$. 
According to theoretical studies of BH binaries in the low/hard state,
the Comptonization continua are likely to be generated in a 
radiatively inefficient accretion flow (RIAF), 
with a predicted spectral slope of $\Gamma \sim 1.4$--1.9 
(e.g., Narayan \& Yi 1994; Esin et al. 1997). 
Therefore, 
the FV continuum is possibly generated in a RIAF-like 
portion of the accretion flow. 
If we multiply 
$L_{\rm th}=6.6\times10^{41}$ erg s$^{-1}$ in 2--10 keV 
by a bolometric correction factor of 10 (e.g., Lusso et al. 2012), 
the critical Eddington ratio below which the RIAF is considered to dominate 
is estimated to be $\sim 0.001$. 
It is slightly smaller than the theoretical value of $\sim 0.01$, 
probably due to uncertainties in the bolometric 
correction factor and the theoretical framework. 

One important characteristic of FV is the neutral absorption 
which is significantly stronger than that to BV. 
According to Narayan \& Yi (1995) and Blandford \& Begelman (1999), 
a RIAF-like disk may drive bipolar outflows because 
of high thermal energy of its hot materials (Ho 2008).  
Such outflows may account for the absorption to FV. 
The case of the 4th observation (\S5) may be explained by 
some changes of the putative absorbers. 
Then, we need a configuration in which BV becomes free from 
the same effect: this is discussed in \S7.5.

If FV corresponds to the low/hard state of BH binaries, 
an analogy to them (e.g., McClintock \& Remillard 2006) predicts 
that radio emission should be detected from NGC 3227 as well, 
when it is in the Faint branch. 
Actually, Kuluta et al. (1995) and Mundell et al. (1995) detected, 
with the VLA and MERLIN observations, respectively, 
a $\sim 40$ pc scale jet emission from NGC 3227. 
These provide a strong support to the prediction. 
However, the radio observation was not covered simultaneously in X-rays, 
so that we cannot tell whether the object was in the Bright or Faint branch 
in X-rays at that time.

%-------------------------------------------------------
\subsection{The neutral Refection component}
%-------------------------------------------------------

The neutral and cold reflection component (3 in \S7.1), 
which is identified as FS and is partly contributing to BS, 
remained almost constant for $\sim1$ month, 
because the normalization of \texttt{pexmon} was confirmed 
unchanged within errors of $\sim10$--20\% through the six observations (Table 4). 
This period,  corresponding to a distance between the central BH and reflectors 
of $\gtrsim 10^4~R_{\rm g}$, suggests that the reflection component 
is mostly generated at an outer part of the accretion disk, or dust torus. 
It is accompanied by a narrow Fe-K$\alpha$ emission line, which has an intensity 
consistent with an Fe abundance of $\sim 1$ Solar.  

It is unclear whether the reflection is generated by BV, or FV, or both,  
because the spectral shape of the reflection is not very sensitive to
the photon index of the incident PL. 
If we assume that the incident PL is due only to FV, 
the reflection fraction of reflectors in the 4th data set would become $f_{\rm ref} = 1.5\pm 0.2$, 
which is significantly larger than the value of $\sim 1$ which is appropriate 
for an infinite plane of the accretion disk. 
On the other hand, the Bright-branch observation, namely the 1st and 3rd ones, 
exhibited $f_{\rm ref} \sim 1$ (Table 4), 
suggesting that both BV and FV may contribute to the generation of the reflection component. 
In the case of NGC 3516 (Noda et al. 2013), both BV and FV are also likely to be 
contributing, as inferred from a solid angle consideration. 
To obtain a firmer answer to this issue, we need a long-term monitoring 
of NGC 3227, because BV is highly variable while the reflection component 
must be averaged over $\sim 1$ month or longer. 

%%%%%%%%%%%%%%%%%%figure12%%%%%%%%%%%%%%%%%%%
\begin{figure*}[t]
\epsscale{1}
\plotone{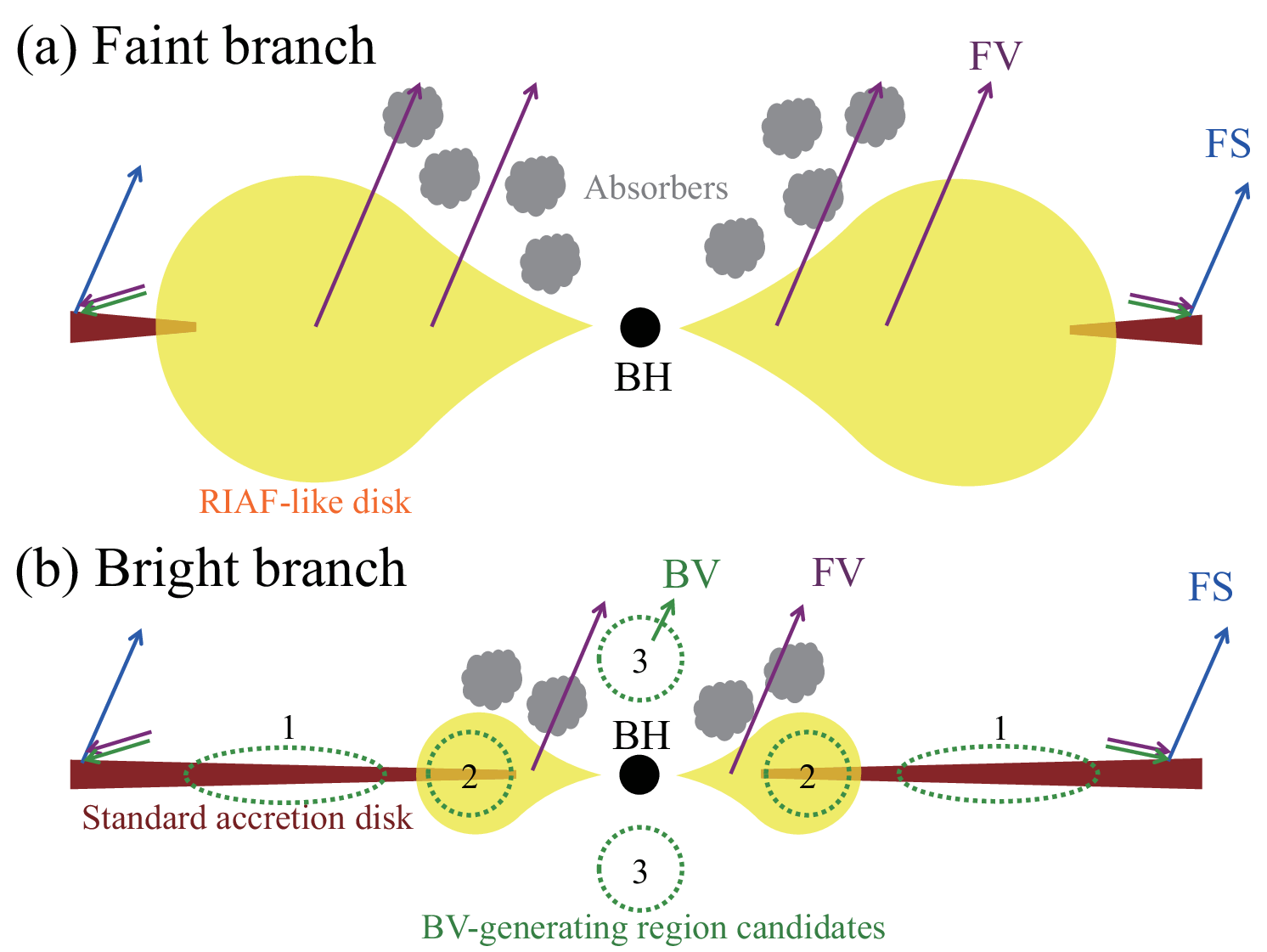}
\caption{Possible geometries of the central engine in NGC 3227, 
in the Faint branch (panel a) and the Bright branch (panel b). In panel (b), 
the candidates for the BV-emitting regions are shown with numbers corresponding to 
those explained in \S7.5. }
\end{figure*}
%%%%%%%%%%%%%%%%%%figure12%%%%%%%%%%%%%%%%%%%%

%-------------------------------------------------------
\subsection{Possible geometry of the central engine in NGC 3227}
%-------------------------------------------------------

Our final task is to discuss geometry of the central engine in NGC 3227, 
trying to explain the present observational results. 
Figure 12 illustrates possible candidates, where we incorporate 
a standard accretion disk and a RIAF-like region, 
of which the relative dominance vary depending on the Eddington ratio. 
In the Faint branch when the luminosity gets lower than $L_{\rm th}$, 
the RIAF-like region is expected to become larger, as shown in Fig. 12(a), 
while the standard accretion disk retracts. 
As a result, the FV component, considered to emerge from the RIAF region, 
dominates the X-ray signal, while BV becomes very weak. 
The FV variations on the timescale of several weeks may directly reflect accretion rate changes. 
The reflection component is still generated by FV, as well as BV if the source 
was relatively bright some times 
%($\gtrsim 1$ month) 
before. 
The FV-generating region presumably gets much larger than 
the distribution of absorbers (outflows), 
making some fraction of the FV component absorbed and the rest not. 
 
As the accretion rate increases and the source enters the Bright branch (Fig. 12b), 
the inner edge of the standard disk becomes closer to the central BH, 
while the RIAF-like region becomes more localized to the innermost part of the disk. 
Therefore, the BV component, presumably originating from a certain location closely 
related to the standard disk, 
is expected to overwhelm the FV component. 
Because the absorbers due to outflows are considered to enshroud the RIAF region, 
only the FV component can be strongly absorbed, while BV not. 
This geometry change form the Faint branch to the Bright branch (panel a to b in Fig. 12)
is considered similar to the transition from the low/hard to the high/soft state of BH binaries, 
and if so, the clear CCP break seen in Fig. 5 and 6
should represent a kind of ``state transition'' in NGC 3227.

Assuming that the BV component requires the presence of an optically-thick disk 
as well as a Comptonizing corona, 
we may specifically identify two candidate regions.
One  (region 1 in Fig. 12b) is small patchy coronal regions located at the
surface of the standard disk (e.g., Reynolds \& Nowak 2003),
which are possibly heated by magnetic processes like Solar coronae 
(e.g., Di Matteo 1998; Machida et al. 2000).
The rapid variability of BV is then attributed to such heating processes 
and physical scales of the regions.
The other, shown as region 2 in Fig. 12(b), is the regions 
where the standard disk and the hot RIAF overlap.
There,  the coronae may be  cooled by rich seed photons from the disk,
so that the BV spectrum steepens
and the absorbing outflows diminish.
Furthermore, even when the accretion rate is constant,
the rapid BV variations can be generated 
by changes of the coronal covering fraction over the disk, 
as proposed by Makishima et al. (2008) 
to explain the fast variations of Cygnus X-1.

There is yet another candidate for the BV-generating site, namely, 
bipolar regions within ~ 25 $R_{\rm g}$ from the central BH (region 3 in Fig. 12b).
There,  some energetic parts of bipolar outflows,
or  ``faulty jets", may be able to produce the BV component
via thermal and/or bulk Comptonization (e.g., Ghisellini et al. 2004).
It remains, however, our future work to further examine 
which of the three candidates are most likely.

We thank all members of the \textit{Suzaku} hardware and software teams and the Science Working Group. 
HN is supported by the Grant-in-Aid for Young Scientists (B) 
(26800095) from the Japan Society for the Promotion of Science (JSPS), 
and the Special Postdoctoral Researchers Program in RIKEN. 
KM  and SS are supported by 
the Grantin-Aid for Scientific Research (A) (23244024) 
from JSPS, and the JSPS Research Fellowship for Young Scientists, respectively.

\end{document}